\documentclass[12pt]{article}

\usepackage{axodraw}
\usepackage{graphicx}
\usepackage{feynarts}

\makeatletter

\def\paragraph{\@startsection{paragraph}{4}{\z@}{+2.00ex plus
 +1ex minus +.2ex}{1.5ex plus .2ex}{\it\normalsize}}

\def\section{\@startsection {section}{1}{\z@}{+3.0ex plus +1ex minus
  +.2ex}{2.3ex plus .2ex}{\normalsize\bf\boldmath}}
\def\subsection{\@startsection{subsection}{2}{\z@}{+2.5ex plus +1ex
minus +.2ex}{1.5ex plus .2ex}{\normalsize\bf\boldmath}}
\def\subsubsection{\@startsection{subsubsection}{3}{\z@}{+3.25ex plus
 +1ex minus +.2ex}{1.5ex plus .2ex}{\normalsize\it}}

\expandafter\ifx\csname mathrm\endcsname\relax\def\mathrm#1{{\rm #1}}\fi

\renewcommand{\theequation}{\thesection.\arabic{equation}}
\newcounter{saveeqn}

\@addtoreset{equation}{section}

\newcount\@tempcntc
\def\@citex[#1]#2{\if@filesw\immediate\write\@auxout{\string\citation{#2}}\fi
  \@tempcnta\z@\@tempcntb\m@ne\def\@citea{}\@cite{\@for\@citeb:=#2\do
    {\@ifundefined
       {b@\@citeb}{\@citeo\@tempcntb\m@ne\@citea
        \def\@citea{,\penalty\@m\ }{\bf ?}\@warning
       {Citation `\@citeb' on page \thepage \space undefined}}%
    {\setbox\z@\hbox{\global\@tempcntc0\csname
b@\@citeb\endcsname\relax}%
     \ifnum\@tempcntc=\z@ \@citeo\@tempcntb\m@ne
       \@citea\def\@citea{,\penalty\@m}
       \hbox{\csname b@\@citeb\endcsname}%
     \else
      \advance\@tempcntb\@ne
      \ifnum\@tempcntb=\@tempcntc
      \else\advance\@tempcntb\m@ne\@citeo
      \@tempcnta\@tempcntc\@tempcntb\@tempcntc\fi\fi}}\@citeo}{#1}}

\def\@citeo{\ifnum\@tempcnta>\@tempcntb\else\@citea
  \def\@citea{,\penalty\@m}%
  \ifnum\@tempcnta=\@tempcntb\the\@tempcnta\else
   {\advance\@tempcnta\@ne\ifnum\@tempcnta=\@tempcntb \else
\def\@citea{--}\fi
    \advance\@tempcnta\m@ne\the\@tempcnta\@citea\the\@tempcntb}\fi\fi}

\def\nl{\nonumber\\}

\newcommand{\lsim}
{\mathrel{\raisebox{-.3em}{$\stackrel{\displaystyle <}{\sim}$}}}
\newcommand{\gsim}
{\mathrel{\raisebox{-.3em}{$\stackrel{\displaystyle >}{\sim}$}}}
\def\asymp#1%
{\mathrel{\raisebox{-.4em}{$\widetilde{\scriptstyle #1}$}}}

\def\Nequal#1%
{\mathrel{\raisebox{-.5em}{$\stackrel{=}{\scriptstyle\rm#1}$}}}
\newcommand{\dsl}[1]{\not \hspace{-0.7mm}#1}
\def\dsl{\mathpalette\make@slash}
\def\make@slash#1#2{\setbox\z@\hbox{$#1#2$}%
  \hbox to 0pt{\hss$#1/$\hss\kern-\wd0}\box0}

\def\beq{\begin{equation}}
\def\eeq{\end{equation}}
\def\beqar{\begin{eqnarray}}
\def\eeqar{\end{eqnarray}}
\def\barr#1{\begin{array}{#1}}
\def\earr{\end{array}}
\def\bfi{\begin{figure}}
\def\efi{\end{figure}}
\def\btab{\begin{table}}
\def\etab{\end{table}}
\def\bce{\begin{center}}
\def\ece{\end{center}}
\def\nn{\nonumber}

\def\text{\textstyle}

\def\arraystretch{1.2}

\def\al{\alpha}
\def\be{\beta}
\def\Ga{\Gamma}
\def\ga{\gamma}
\def\de{\delta}
\def\De{\Delta}
\def\eps{\epsilon}

\def\la{\lambda}

\def\si{\sigma}

\def\refeq#1{\mbox{(\ref{#1})}}

\def\reffi#1{\mbox{Figure~\ref{#1}}}

\def\refta#1{\mbox{Table~\ref{#1}}}

\def\refse#1{\mbox{Section~\ref{#1}}}

\def\citere#1{\mbox{Ref.~\cite{#1}}}
\def\citeres#1{\mbox{Refs.~\cite{#1}}}

\newcommand{\TeV}{\unskip\,\mathrm{TeV}}
\newcommand{\GeV}{\unskip\,\mathrm{GeV}}
\newcommand{\MeV}{\unskip\,\mathrm{MeV}}

\newcommand{\fb}{\unskip\,\mathrm{fb}}

\newcommand{\ri}{{\mathrm{i}}}
\newcommand{\rd}{{\mathrm{d}}}
\newcommand{\rE}{{\mathrm{E}}}

\newcommand{\Oa}{\mathswitch{{\cal{O}}(\alpha)}}

\newcommand{\Oaaa}{\mathswitch{{\cal{O}}(\alpha^3)}}

\newcommand{\M}{{\cal{M}}}

\def\mathswitchr#1{\relax\ifmmode{\mathrm{#1}}\else$\mathrm{#1}$\fi}
\newcommand{\Pf}{\mathswitch  f}
\newcommand{\Pfbar}{\mathswitch{\bar f}}

\newcommand{\PW}{\mathswitchr W}
\newcommand{\Pw}{\mathswitchr w}
\newcommand{\PZ}{\mathswitchr Z}

\newcommand{\PH}{\mathswitchr H}
\newcommand{\Pe}{\mathswitchr e}
\newcommand{\Pp}{\mathswitchr p}
\newcommand{\Pne}{\mathswitch \nu_{\mathrm{e}}}
\newcommand{\Pnebar}{\mathswitch \bar\nu_{\mathrm{e}}}
\newcommand{\Pd}{\mathswitchr d}

\newcommand{\Pu}{\mathswitchr u}

\newcommand{\Ps}{\mathswitchr s}

\newcommand{\Pc}{\mathswitchr c}

\newcommand{\Pb}{\mathswitchr b}

\newcommand{\Pt}{\mathswitchr t}

\newcommand{\Pepm}{\mathswitchr {e^\pm}}
\newcommand{\Pep}{\mathswitchr {e^+}}
\newcommand{\Pem}{\mathswitchr {e^-}}
\newcommand{\PWp}{\mathswitchr {W^+}}
\newcommand{\PWm}{\mathswitchr {W^-}}

\def\mathswitch#1{\relax\ifmmode#1\else$#1$\fi}

\newcommand{\MW}{\mathswitch {M_\PW}}

\newcommand{\MZ}{\mathswitch {M_\PZ}}
\newcommand{\MH}{\mathswitch {M_\PH}}
\newcommand{\Me}{\mathswitch {m_\Pe}}

\newcommand{\Md}{\mathswitch {m_\Pd}}
\newcommand{\Mu}{\mathswitch {m_\Pu}}
\newcommand{\Ms}{\mathswitch {m_\Ps}}
\newcommand{\Mc}{\mathswitch {m_\Pc}}
\newcommand{\Mb}{\mathswitch {m_\Pb}}
\newcommand{\Mt}{\mathswitch {m_\Pt}}

\newcommand{\GZ}{\Gamma_{\PZ}}

\newcommand{\sw}{\mathswitch {s_\Pw}}
\newcommand{\cw}{\mathswitch {c_\Pw}}

\newcommand{\GF}{\mathswitch {G_\mu}}

\def\solid{\raise.9mm\hbox{\protect\rule{1.1cm}{.2mm}}}
\def\dash{\raise.9mm\hbox{\protect\rule{2mm}{.2mm}}\hspace*{1mm}}

\def\ie{i.e.\ }

\def\sub{{\mathrm{sub}}}
\def\gsub{g^{(\sub)}}
\def\Gsub{G^{(\sub)}}
\def\cGsub{{\cal G}^{(\sub)}}
\def\cGcoll{{\cal G}^{(\coll)}}

\newcommand{\LEP}{{\mathrm{LEP}}}

\newcommand{\tree}{{\mathrm{tree}}}

\newcommand{\IBA}{{\mathrm{IBA}}}

\newcommand{\WW}{{\mathrm{WW}}}
\newcommand{\ZH}{{\mathrm{ZH}}}
\newcommand{\virt}{{\mathrm{virt}}}
\newcommand{\soft}{{\mathrm{soft}}}
\newcommand{\coll}{{\mathrm{coll}}}
\newcommand{\finite}{{\mathrm{finite}}}

\def\Li{\mathop{\mathrm{Li}_2}\nolimits}

\def\Re{\mathop{\mathrm{Re}}\nolimits}
\def\Im{\mathop{\mathrm{Im}}\nolimits}

\newcommand{\eennh}{\Pep\Pem\to\nu\bar\nu\PH}
\newcommand{\eeneneh}{\Pep\Pem\to\Pne\Pnebar\PH}
\newcommand{\eenmnmh}{\Pep\Pem\to\nu_\mu\bar\nu_\mu\PH}

\def\draftdate{\relax}
\def\mda{\relax}
\def\mua{\relax}
\def\mla{\relax}
\def\draft{
\def\thtystars{******************************}
\def\sixtystars{\thtystars\thtystars}
\typeout{}
\typeout{\sixtystars**}
\typeout{* Draft mode!
         For final version remove \protect\draft\space in source file *}
\typeout{\sixtystars**}
\typeout{}
\def\draftdate{\today}
\def\mua{\marginpar[\boldmath\hfil$\uparrow$]%
                   {\boldmath$\uparrow$\hfil}%
                    \typeout{marginpar: $\uparrow$}\ignorespaces}
\def\mda{\marginpar[\boldmath\hfil$\downarrow$]%
                   {\boldmath$\downarrow$\hfil}%
                    \typeout{marginpar: $\downarrow$}\ignorespaces}
\def\mla{\marginpar[\boldmath\hfil$\rightarrow$]%
                   {\boldmath$\leftarrow $\hfil}%
                    \typeout{marginpar: $\leftrightarrow$}\ignorespaces}
\def\Mua{\marginpar[\boldmath\hfil$\Uparrow$]%
                   {\boldmath$\Uparrow$\hfil}%
                    \typeout{marginpar: $\uparrow$}\ignorespaces}
\def\Mda{\marginpar[\boldmath\hfil$\Downarrow$]%
                   {\boldmath$\Downarrow$\hfil}%
                    \typeout{marginpar: $\downarrow$}\ignorespaces}
\def\Mla{\marginpar[\boldmath\hfil$\Rightarrow$]%
                   {\boldmath$\Leftarrow $\hfil}%
                    \typeout{marginpar: $\leftrightarrow$}\ignorespaces}
\overfullrule 5pt
\oddsidemargin -15mm
\marginparwidth 29mm
}

\def\stars{\strut\leaders\hbox{*}\hfill\strut}
\def\starline{\hfil\strut\hfil\hbox to \textwidth {\stars}\hfil}

\begin{document}
\thispagestyle{empty}
\def\thefootnote{\fnsymbol{footnote}}
\setcounter{footnote}{1}
\null
\draftdate\hfill KA-TP-03-2003\\
\strut\hfill MPI-PhT/2003-08 \\
\strut\hfill PSI-PR-03-05\\
\strut\hfill hep-ph/0302198
\vfill
\begin{center}
{\Large \bf\boldmath
Electroweak radiative corrections to $\Pep\Pem\to \nu\bar\nu\PH$%
\par} \vskip 2.5em
\vspace{1cm}

{\large
{\sc A.\ Denner$^1$, S.\ Dittmaier$^2$, M. Roth$^3$ and 
M.~M.~Weber$^1$} } \\[1cm]
$^1$ {\it Paul-Scherrer-Institut, W\"urenlingen und Villigen\\
CH-5232 Villigen PSI, Switzerland} \\[0.5cm]
$^2$ {\it Max-Planck-Institut f\"ur Physik 
(Werner-Heisenberg-Institut) \\
D-80805 M\"unchen, Germany}
\\[0.5cm]
$^3$ {\it Institut f\"ur Theoretische Physik, Universit\"at Karlsruhe \\
D-76128 Karslruhe, Germany}
\par \vskip 1em
\end{center}\par
\vskip 2cm {\bf Abstract:} \par The complete electroweak ${\cal
  O}(\alpha)$ radiative corrections to the Higgs-boson production
processes $\Pep\Pem\to \nu_l\bar\nu_l\PH$ ($l=\Pe,\mu,\tau$) are
calculated in the electroweak Standard Model.  For
$\Pep\Pem\to\nu_\Pe\bar\nu_\Pe\PH$, where $\PZ\PH$ production and
W-boson fusion contribute, both production channels are added
coherently. The calculation of the corrections is described in some
detail including, in particular, the treatment of the Z-boson
resonance in the $\PZ\PH$-production channel.  The discussion of
numerical results focusses on the total cross section as well as on
angular and energy distributions of the Higgs boson. In the
$\GF$-scheme, the bulk of the corrections is due to initial-state
radiation.  The corrections turn out to reduce the total cross section
by $\sim 10\%$ for high energies, where the W-boson fusion dominates.
In this region, the corrections depend only weakly on the energy and
the production angle of the Higgs boson.  Based on an analysis of the
leading universal corrections, a simple improved Born approximation is
introduced.  This approximation describes the corrected cross section
within about 3\%.
\par
\vskip 1cm
\noindent
February 2003
\null
\setcounter{page}{0}
\clearpage
\def\thefootnote{\arabic{footnote}}
\setcounter{footnote}{0}

\section{Introduction}
\label{se:intro}

One of the most important open problems of particle physics is the
understanding of the mechanism of electroweak symmetry breaking. In
the electroweak Standard Model (SM) it is provided by the Higgs
mechanism, leading to the prediction of a physical scalar particle,
the Higgs boson. The investigation of the mechanism of electroweak
symmetry breaking and, in particular, of the Higgs boson, will be one
of the main concerns at the Large Hadron Collider (LHC) at CERN. The
LHC experiments ATLAS \cite{ATLAS} and CMS \cite{CMS} are sensitive to
the SM Higgs boson over the whole mass range from the present lower
experimental limit of $114.4\GeV$ \cite{:2001xw} up to $1\TeV$ and
will discover the Higgs boson, if it exists and has no particularly
exotic properties. Moreover, these experiments will determine various
properties of the Higgs boson, such as its mass, branching ratios, and
ratios of its coupling constants.

However, the complete profile of the Higgs boson can only be studied
in the clean environment of an electron--positron linear collider
\cite{Accomando:1998wt,Aguilar-Saavedra:2001rg,Abe:2001gc,Abe:2001wn}.
In $\Pep\Pem$ annihilation there are two main production mechanisms
for the SM Higgs boson. In the Higgs-strahlung process,
$\Pep\Pem\to\PZ\PH$, a virtual $\PZ$ boson decays into a $\PZ$ boson
and a Higgs boson.  The corresponding cross section rises sharply at
threshold to a maximum of a few tens of GeV above $\MZ+\MH$ and then
falls off as $s^{-1}$, where $\sqrt{s}$ is the centre-of-mass (CM)
energy of the $\Pep\Pem$ system.  In the W-boson-fusion process,
$\Pep\Pem\to\Pne\Pnebar\PH$, the incoming $\Pep$ and $\Pem$ each emit
a virtual W~boson which fuse into a Higgs boson. The cross section of
the W-boson-fusion process grows as $\ln s$ and thus is the dominant
production mechanism for $\sqrt{s}\gg\MH$. The cross section for the
similar $\PZ$-boson-fusion process, $\Pep\Pem\to\Pep\Pem\PH$, is about
one order of magnitude smaller.

At a linear $\Pep\Pem$ collider with a CM energy of $500\GeV$ and an
integrated luminosity of $500\fb^{-1}$, of the order of $10^4$ Higgs
bosons can be produced per year \cite{Aguilar-Saavedra:2001rg}. This
allows to measure the Higgs-production cross sections and thus the
Higgs--gauge-boson couplings at the level of a few per cent.
Consequently, adequate theoretical predictions have to take into
account radiative corrections and the relevant effects of the finite
decay width of the $\PZ$ and Higgs bosons.  For a heavy Higgs, which
decays mainly into W-boson pairs, finite-width effects have been
investigated in \citere{Accomando:1997gj,Dittmaier:2002ap}.  On the
other hand, the effects of the finite width of the Higgs boson can be
neglected if the Higgs boson is light, \ie if its width is small.

The Higgs-strahlung process has been investigated in lowest order in
\citere{Ellis:1975ap}. Including the \PZ-boson decay, the process
$\Pep\Pem\to\PZ^*\PH\to\Pf\Pfbar\PH$ has been studied by Bjorken
\cite{Bj76}.  For the process $\Pep\Pem\to\PZ\PH$, the $\Oa$
electroweak corrections have been calculated in the soft-photon
approximation in
\citeres{Fleischer:1982af,Kniehl:1991hk,Denner:1992bc}, and a Monte
Carlo algorithm for the calculation of the real photonic corrections
to this process was described in \citere{Berends:dw}. A compact
analytical formula for the electromagnetic corrections to the total
cross section can be found in \citere{Kniehl:1991hk}.

The vector-boson-fusion processes have been investigated in lowest
order in \citere{Jones:1979bq}.  The electroweak corrections to
$\eennh$ have attracted a lot of interest recently.  Analytical
results for the one-loop corrections to this process have been
obtained in \citere{Jegerlehner:2002es} as {\sl MAPLE} output, but a
numerical evaluation of these results is not yet available. A first
complete calculation of the $\Oa$ electroweak corrections to $\eennh$
in the SM has been performed in \citere{Belanger:2002me}.  The
contributions of fermion and sfermion loops in the Minimal
Supersymmetric Standard Model have been evaluated in
\citeres{Eberl:2002xd,Hahn:2002gm} with seemingly differing results.
We have performed a completely independent calculation of the $\Oa$
electroweak corrections to the complete process $\eennh$ in the SM.
First results of our calculation have already been presented in
\citere{Denner:2003yg}. There, we have successfully compared our
results for the total cross section with those of
\citeres{Belanger:2002me,Eberl:2002xd,Hahn:2002gm} and pointed out
that the differences between \citeres{Eberl:2002xd} and
\cite{Hahn:2002gm} are due to different renormalization schemes and
input parameters.

In this paper, we present the details of our calculation of the $\Oa$
electroweak corrections to the processes $\eeneneh$,
$\nu_\mu\bar\nu_\mu\PH$, and $\nu_\tau\bar\nu_\tau\PH$. While the
first process gets contributions from W~fusion and Higgs-strahlung,
the final states with $\mu$ or $\tau$ neutrinos receive contributions
only from Higgs-strahlung off Z~bosons.  Single hard-photon radiation
is included using the complete matrix element, and higher-order ISR
corrections are taken into account in the leading-logarithmic
approximation.  The finite Z-boson decay width is introduced in the
constant-width scheme and in a gauge-invariant scheme that is
based on a factorization of the Z~resonance in the gauge-invariant set
of diagrams related to the (neutral-current) Higgs-strahlung process.
 
The paper is organized as follows: in \refse{se:calcrcs} the
calculation of the virtual, real, and (higher-order)
leading-logarithmic corrections is described, and the leading
universal corrections are discussed.  Section \ref{se:numres} contains
a discussion of numerical results. The paper is summarized in
\refse{se:sum}. Further useful information on the calculation is
collected in the Appendix.

\section{Calculation of radiative corrections}
\label{se:calcrcs}

\subsection{Conventions and lowest-order cross section}
\label{se:convs}

We consider the processes
\beq
\Pem(p_1,\sigma_1) + \Pep(p_2,\sigma_2) \;\longrightarrow\;
\nu_l(k_1) + \bar\nu_l(k_2) + \PH(k_3), \qquad l=\Pe,\mu,\tau,
\label{eq:eennh}
\eeq
where the momenta $p_i$, $k_j$ of all particles and the electron
helicities $\sigma_i$ are given in parentheses.  The helicities take
the values $\sigma_i=\pm1/2$, but we often use only the sign to
indicate the helicity.  The electron mass is neglected whenever
possible, i.e.\ it is kept finite only in the mass-singular logarithms
related to initial-state radiation. This implies that the lowest-order
and one-loop amplitudes vanish unless $\si_1=-\si_2$. Therefore, we
define $\si=\si_1=-\si_2$.  The particle momenta obey the mass-shell
conditions $p_1^2=p_2^2=k_1^2=k_2^2=0$ and $k_3^2=\MH^2$. For later
use, the following set of kinematical invariants is defined:
\beqar
s &=& (p_1+p_2)^2, \nn\\
s_{ij} &=& (k_i+k_j)^2, \qquad i,j=1,2,3, \nn\\
t_{ij} &=& (p_i-k_j)^2, \qquad i=1,2, \quad j=1,2,3.
\eeqar

In lowest order the processes \refeq{eq:eennh} proceed via the
diagrams shown in \reffi{fi:borndiags}. More precisely, only for
electron neutrinos in the final state ($l=\Pe$) both the
$\PZ\PH$-production and $\PW\PW$-fusion diagrams contribute, while for
$\mu$ and $\tau$ neutrinos merely the former exists.
\begin{figure}                                                    
\centerline{\footnotesize \input{paper-tree-a} \hspace{3em}
                    \input{paper-tree-b} } 
\caption{Lowest-order diagrams for $\Pem\Pep\to\nu\bar\nu\PH$} 
\label{fi:borndiags} 
\end{figure}        

In the calculation of the tree-level amplitude $\M^\si_0$ and of the
one-loop amplitude $\M^\si_1$, which is described in the next section,
we separate the fermion spinor chains by defining standard matrix
elements (SME).  To introduce a compact notation for the SME, the
tensors
\beqar
\Gamma^{\Pe\Pe,\si}_{\{\al,\al\be\ga\}} &=&
\bar v_{\Pep}(p_2)\left\{\ga_\al,\ga_\al\ga_\be\ga_\ga\right\}
\omega_\si u_{\Pem}(p_1),
\nn\\
\Gamma^{\nu\nu}_{\{\al,\al\be\ga\}} &=&
\bar u_{\nu_l}(k_1)\left\{\ga_\al,\ga_\al\ga_\be\ga_\ga\right\}
\omega_- v_{\bar\nu_l}(k_2),
\nn\\
\Gamma^{\nu\Pe}_{\{\al,\al\be\ga\}} &=&
\bar u_{\nu_l}(k_1)\left\{\ga_\al,\ga_\al\ga_\be\ga_\ga\right\}
\omega_- u_{\Pem}(p_1),
\nn\\
\Gamma^{\Pe\nu}_{\{\al,\al\be\ga\}} &=&
\bar v_{\Pep}(p_2)\left\{\ga_\al,\ga_\al\ga_\be\ga_\ga\right\}
\omega_- v_{\bar\nu_l}(k_2)
\eeqar
are defined with obvious notations for the Dirac spinors $\bar
v_{\Pep}(p_2)$, etc., and $\omega_\pm=(1\pm\gamma_5)/2$ denote the
right- and left-handed chirality projectors.  Here and in the
following, each entry in the set within curly brackets refers to a
single object, i.e.\ from the first line in the equation above we have
$\Gamma^{\Pe\Pe,\si}_{\al} = \bar v_{\Pep}(p_2)\ga_\al \omega_\si
u_{\Pem}(p_1)$, etc.  Furthermore, symbols like $\Gamma_p$ are used as
shorthand for the contraction $\Gamma_\mu\, p^\mu$.  For the
$\PZ\PH$-production channel we define the 26 SME
\newcommand{\Msme}{\hat\M}
\beq
\def\arraystretch{1.5}
\begin{array}[b]{rclcrcl}
\Msme^{\ZH,\si}_{\{1,2\}} &=&
\Ga^{\Pe\Pe,\si}_{\al} \; \Ga^{\nu\nu,\{\al,\al{p}_1{p}_2\}}, 
& \qquad &
\Msme^{\ZH,\si}_{\{3,4\}} &=& 
\Ga^{\Pe\Pe,\si}_{\al k_1 k_2} \; \Ga^{\nu\nu,\{\al,\al{p}_1{p}_2\}}, 
\\
\Msme^{\ZH,\si}_{\{5,6\}} &=& 
\Ga^{\Pe\Pe,\si}_{k_1} \; \Ga^{\nu\nu,\{{p}_1,{p}_2\}}, 
& \qquad &
\Msme^{\ZH,\si}_{\{7,8\}} &=& 
\Ga^{\Pe\Pe,\si}_{k_2} \; \Ga^{\nu\nu,\{{p}_1,{p}_2\}}, 
\\
\Msme^{\ZH,\si}_{\{9,10\}} &=& 
\Ga^{\Pe\Pe,\si}_{\al\be k_1} \; \Ga^{\nu\nu,\{\al\be {p}_1,\al\be {p}_2\}}, 
& \qquad &
\Msme^{\ZH,\si}_{\{11,12\}} &=& 
\Ga^{\Pe\Pe,\si}_{\al\be k_2} \; \Ga^{\nu\nu,\{\al\be {p}_1,\al\be {p}_2\}}, 
\\
\Msme^{\ZH,\si}_{13} &=& 
\Ga^{\Pe\Pe,\si}_{\al\be\ga} \; \Ga^{\nu\nu,\al\be\ga}. 
&&&&
\end{array}
\eeq
For the $\PW\PW$-fusion channel we introduce the following set of 13
SME,
\beq
\def\arraystretch{1.5}
\begin{array}[b]{rclcrcl}
\Msme^{\WW}_{\{1,2\}} &=&
\Ga^{\nu\Pe}_{\al} \; \Ga^{\Pe\nu,\{\al,\al{k}_1{p}_1\}}, 
& \qquad &
\Msme^{\WW}_{\{3,4\}} &=& 
\Ga^{\nu\Pe}_{\al k_2 p_2} \; \Ga^{\Pe\nu,\{\al,\al{k}_1{p}_1\}}, 
\\
\Msme^{\WW}_{\{5,6\}} &=& 
\Ga^{\nu\Pe}_{p_2} \; \Ga^{\Pe\nu,\{{p}_1,{k}_1\}}, 
& \qquad &
\Msme^{\WW}_{\{7,8\}} &=& 
\Ga^{\nu\Pe}_{k_2} \; \Ga^{\Pe\nu,\{{p}_1,{k}_1\}}, 
\\
\Msme^{\WW}_{\{9,10\}} &=& 
\Ga^{\nu\Pe}_{\al\be p_2} \; \Ga^{\Pe\nu,\{\al\be {p}_1,\al\be {k}_1\}}, 
& \qquad &
\Msme^{\WW}_{\{11,12\}} &=& 
\Ga^{\nu\Pe}_{\al\be k_2} \; \Ga^{\Pe\nu,\{\al\be {p}_1,\al\be {k}_1\}}, 
\\
\Msme^{\WW}_{13} &=& 
\Ga^{\nu\Pe}_{\al\be\ga} \; \Ga^{\Pe\nu,\al\be\ga}. 
&&&&
\end{array}
\eeq

The tree-level and one-loop amplitudes can be expanded in terms of 
linear combinations of SME,
\beq
\M^\si_n = \sum_{i=1}^{13} F^{\ZH,\si}_{n,i} \Msme^{\ZH,\si}_i \;+\;
\de_{\si-} \sum_{i=1}^{13} F^{\WW}_{n,i} \Msme^{\WW}_i, \qquad n=0,1,
\eeq
with Lorentz-invariant functions $F^{\ZH,\si}_{n,i}$ and
$F^{\WW}_{n,i}$.  This decomposition is unique in $D$ dimensions, \ie
if only the Dirac equation for the spinors is used. The number of SME
can, however, be reduced further by exploiting the four-dimensionality
of space--time, which implies relations among the SME given above.  In
fact, it is possible to express the set of all 39 SME in terms of two
independent SME.  We list the expressions of the two independent SME,
which can be identified with $\Msme^{\ZH,\pm}_1$, and the relations
among the SME in the Appendix.

The lowest-order amplitude reads
\beq
\M^\si_0 = \M^{\ZH,\si}_0 + \de_{\si-} \M^{\WW}_0,
\label{eq:M0}
\eeq
where
\beqar
\M^{\ZH,\si}_0 &=&
-\frac{e^3\MW}{2\cw^3\sw^2} \frac{1}{(s-\MZ^2)(s_{12}-\MZ^2+\ri\MZ\GZ)}
\, g^\si_{\PZ\Pe}\Msme^{\ZH,\si}_1,
\label{eq:MZH0}
\\
\M^{\WW}_0 &=& 
\delta_{l\Pe} \, \frac{e^3\MW}{2\sw^3} 
\frac{1}{(t_{11}-\MW^2)(t_{22}-\MW^2)} \, \Msme^{\WW}_1
\label{eq:MWW0}
\eeqar
with the chiral couplings $g^\si_{\PZ\Pe}$ of the
electron to the Z~boson,
\beq
g^\si_{\PZ\Pe} = \frac{\sw}{\cw} - \frac{\de_{\si -}}{2\cw\sw}.
\eeq
The sine and cosine of the weak mixing angle are fixed by
\beq\label{eq:defcw}
\cw^2 = 1-\sw^2 = \frac{\MW^2}{\MZ^2}.
\eeq
For $s_{12}\to\MZ^2$, the lowest-order amplitude develops a resonance
corresponding to real $\PZ\PH$ production with the subsequent
$\PZ\to\nu_l\bar\nu_l$ decay. We, therefore, have included a finite
Z-boson width $\GZ$ in the denominator, which results from a (partial)
Dyson summation of the corresponding propagator. This procedure is
discussed in more detail in the context of the virtual radiative
corrections in the next section.

Finally, the lowest-order cross section reads
\beq
\sigma_0 = \frac{1}{2s} \,
\int\rd\Phi_3 \sum_{\si=\pm\frac{1}{2}}
\frac{1}{4}(1+2P_-\si)(1-2P_+\si) \,
|\M_0^\si|^2,
\label{eq:sigma0}
\eeq
where $P_\pm$ are the degrees of polarization of the $\Pe^\pm$ beams and
the phase-space integral is defined by
\beq
\int \rd\Phi_3 =
\left( \prod_{i=1}^3 \int\frac{\rd^3 {\bf k}_i}{(2\pi)^3 2k_i^0} \right)\,
(2\pi)^4 \delta\Biggl(p_1+p_2-\sum_{j=1}^3 k_j\Biggr).
\label{eq:dG3}
\eeq

\subsection{Virtual corrections}
\label{se:vrcs}

\subsubsection{Survey of one-loop diagrams}

The virtual corrections can be classified into self-energy, vertex,
box, and pentagon corrections. The generic contributions of the
different vertex functions are shown in \reffi{fi:gendiagrams}.
\begin{figure}
\centerline{\footnotesize  \input{paper-vertex}}
\caption{Contributions of different vertex functions to $\eennh$}
\label{fi:gendiagrams}
\end{figure}
The first three lines contain those diagrams that contribute to all
$\nu_l\bar\nu_l\PH$ final states, whereas the diagrams in the last
three lines contribute only to $\Pep\Pem\to\nu_\Pe\bar\nu_\Pe\PH$.

The complete set of pentagon diagrams is shown in \reffi{fi:pentagons}.
\begin{figure}
\centerline{\footnotesize  \input{paper-pentagon}}
\caption{Pentagon diagrams for $\eennh$}
\label{fi:pentagons}
\end{figure}
The last eight diagrams contribute only for the
$\nu_\Pe\bar\nu_\Pe\PH$ final state.
The $\nu_l\bar{\nu}_l\PZ\PH$ and $\nu_l\bar{\nu}_l\ga\PH$ box
diagrams are depicted in \reffi{fi:nnZH}, the $\Pep\Pem\PZ\PH$ box
diagrams in \reffi{fi:eeZH}, and the $\Pem\Pnebar\PWp\PH$ box diagrams
in \reffi{fi:enWH}.
\begin{figure}
\centerline{\footnotesize  \input{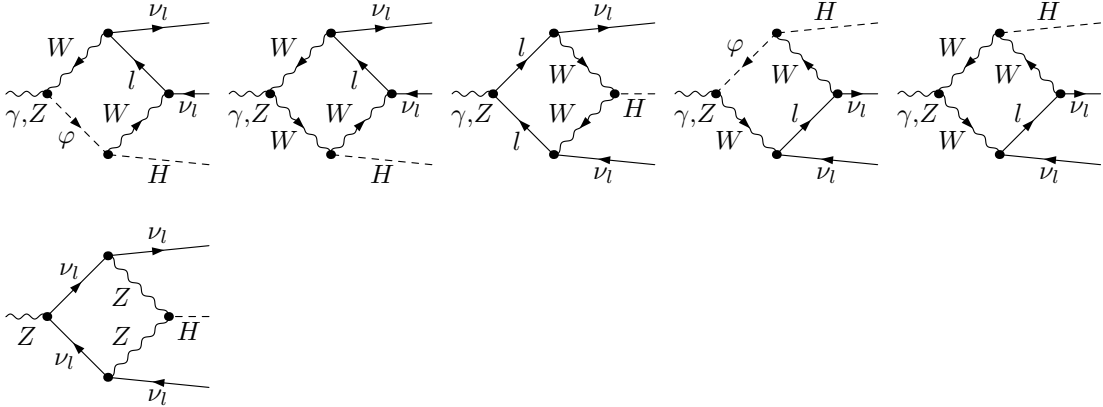}}
\caption{Diagrams for the $\nu_l \bar{\nu}_l\PZ\PH$ and
$\nu_l \bar{\nu}_l\gamma\PH$ vertex functions}
\label{fi:nnZH}
\end{figure}
\begin{figure}
\centerline{\footnotesize  \input{paper-eeZH}}
\caption{Diagrams for the $\Pep\Pem\PZ\PH$ vertex function}
\label{fi:eeZH}
\end{figure}
\begin{figure}
\centerline{\footnotesize  \input{paper-enWH}}
\caption{Diagrams for the $\Pem\Pnebar\PWp\PH$ vertex function}
\label{fi:enWH}
\end{figure}
Those for the $\Pep\Pne\PWm\PH$ box diagrams can be obtained by charge conjugation.

\begin{figure}
\centerline{\footnotesize  \input{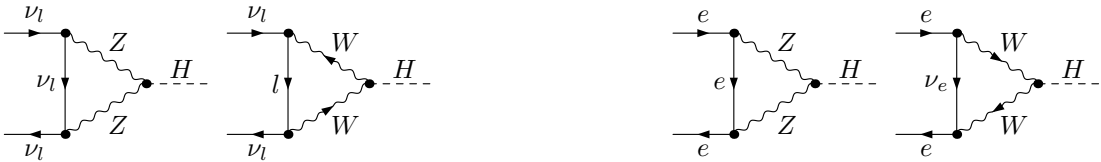}}
\caption{Diagrams for the $\nu_l \bar{\nu}_l\PH$ and 
$\Pep\Pem\PH$ vertex functions}
\label{fi:ffH}
\end{figure}

The diagrams for the $\nu_l \bar{\nu}_l\PH$ and $\Pep\Pem\PH$ vertex
functions are listed in \reffi{fi:ffH}.
Figure \ref{fi:WWH} shows the Feynman diagrams for the $\PWp\PWm\PH$
vertex function and \reffi{fi:ZZH} those for the $\PZ\PZ\PH$ and
$\gamma\PZ\PH$ vertex functions.
\begin{figure}
\centerline{\footnotesize  \input{paper-WWH}}
\caption{Diagrams for the $\PWp\PWm\PH$ vertex function}
\label{fi:WWH}
\end{figure}
\begin{figure}
\centerline{\footnotesize  \input{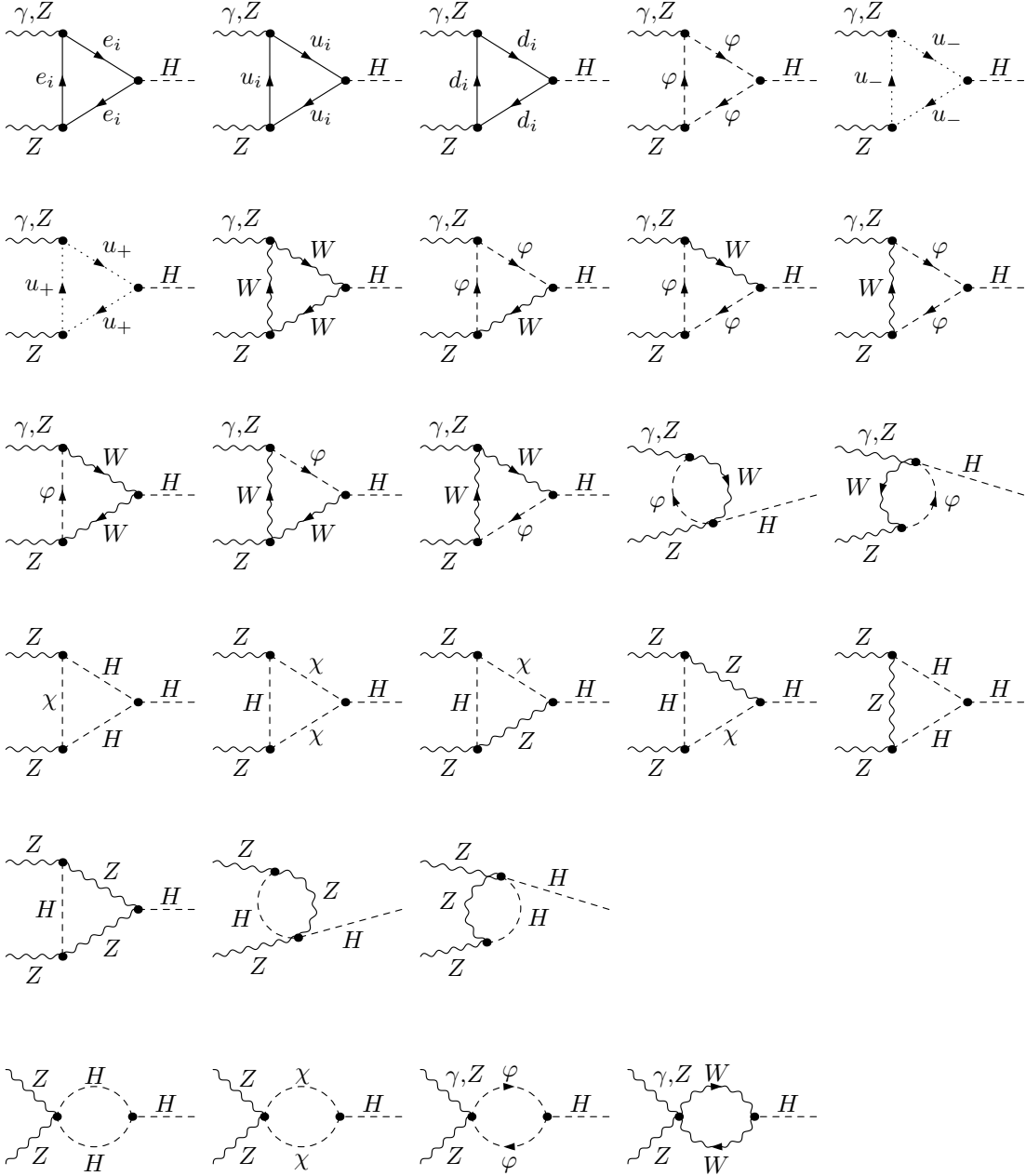}}
\caption{Diagrams for the $\PZ\PZ\PH$ and $\gamma \PZ\PH$ vertex functions}
\label{fi:ZZH}
\end{figure}
Note that in \reffi{fi:ZZH} those diagrams that are obtained from the
diagrams in the first three lines of this figure by reversing the
charge flow in the loop have been suppressed. Most of the diagrams for
the self-energies and the $\nu_l \bar{\nu}_l\PZ$, $\Pep\Pem\PZ$, and
$\Pepm{\stackrel{(-)}{\Pne}}\PW$ vertex functions can be found in
\citere{Hollik:1988ii}.

All pentagon and box diagrams are UV finite, and also the
$\Pep\Pem\PH$ and $\nu_l \bar{\nu}_l\PH$ vertex functions are finite
since we neglect the electron mass everywhere apart from the
mass-singular logarithms.  For the other vertex functions,
$\ga\PZ\PH$, $\PZ\PZ\PH$, $\nu_l \bar{\nu}_l\PZ$, $\Pep\Pem\PZ$,
$\PWp\PWm\PH$, $\Pepm{\stackrel{(-)}{\Pne}}\PW$, and for the $\PZ\PZ$,
$\ga\PZ$, and $\PW\PW$ self energies the corresponding counterterm
diagrams have to be included.

\subsubsection{Calculational framework}

The actual calculation of the one-loop diagrams has been carried out
in the 't~Hooft--Feynman gauge using standard techniques. The Feynman
graphs have been generated with {\sl Feyn\-Arts}
\cite{Kublbeck:1990xc} and are evaluated in two completely independent
ways, leading to two independent computer codes.  The results of the
two codes are in good numerical agreement (i.e.\ within about 12
digits for non-exceptional phase-space points).  Apart from the
5-point integrals, in both calculations the tensor coefficients of the
one-loop integrals are algebraically reduced to scalar integrals with
the Passarino--Veltman algorithm \cite{Passarino:1979jh} at the
numerical level.  The scalar integrals are evaluated using the methods
and results of
\citeres{'tHooft:1979xw,Beenakker:1990jr,Denner:1993kt}, where
ultraviolet divergences are regulated dimensionally and IR divergences
with an infinitesimal photon mass $m_\gamma$.  The renormalization is
carried out in the on-shell renormalization scheme, as e.g.\ described
in \citere{Denner:1993kt}.

In the first calculation, the Feynman graphs are generated with {\sl
  Feyn\-Arts} version 1.0 \cite{Kublbeck:1990xc}.  With the help of
{\sl Mathematica} routines the amplitudes are expressed in terms of
SME and coefficients of tensor integrals. The output is processed into
a {\sl Fortran} program for the numerical evaluation.  For the
evaluation of the tensor 5-point function two approaches have been
followed: the usual Passarino--Veltman reduction and the direct
reduction to 4-point integrals of \citere{Denner:2002ii}.  The results
based on the Passarino--Veltman algorithm become numerically unstable
at the phase-space boundary and could only be rescued by a careful
extrapolation out of the numerically safe inner phase-space domains,
as described in Section 2.2.4 of \citere{Beenakker:2002nc}. The
instabilities are due to the occurrence of inverse Gram determinants
in the recursive tensor reduction. The direct reduction of
\citere{Denner:2002ii} avoids such inverse Gram determinants,
rendering the results of this approach well behaved near the
phase-space boundary.

The calculation of the virtual corrections has been repeated using the
background-field method \cite{Denner:1994xt}, where the individual
contributions from self-energy, vertex, box, and pentagon corrections
differ from their counterparts in the conventional formalism.  The
total one-loop corrections of the conventional and of the
background-field approach were found to be in perfect numerical
agreement.

The second calculation has been made using {\sl FeynArts} version~3
\cite{Hahn:2000kx} for the generation and {\sl FormCalc}
\cite{Hahn:1998yk} for the evaluation of the amplitudes. The
analytical results of {\sl FormCalc} in terms of SME and their
coefficients were translated into {\sl C++} code, and the interference
of the one-loop with the lowest-order amplitude calculated
numerically.  However, in the case of the pentagon diagrams this
interference was calculated analytically using {\sl FeynCalc}
\cite{Mertig:an}.  After evaluation of the fermion traces, the loop
momenta appear in the numerator only in scalar products and can be
cancelled against propagator denominators.  In this way, only scalar
5-point functions remain, thereby avoiding inverse Gram determinants
and the related numerical instabilities.

Finally, the contribution of the virtual corrections to the cross
section is given by
\beq
\de\sigma_{\virt} = \frac{1}{2s} \,
\int\rd\Phi_3 \sum_{\si=\pm\frac{1}{2}}
\frac{1}{4}(1+2P_-\si)(1-2P_+\si) \,
2\Re\left\{\M_1^\si\left(\M_0^\si\right)^*\right\}.
\label{eq:sigmaV}
\eeq

\subsubsection{Treatment of the Z-boson resonance}
\label{se:resonance}

At tree level the introduction of the Z-boson decay width in the
lowest-order matrix element $\M^\si_0$ of Eq.~\refeq{eq:M0} did not
pose any problem with gauge invariance or double-counting of
higher-order effects, since the gauge-invariant $\PZ\PH$-production
part before Dyson summation, $\M_0^{\ZH,\si}(\GZ=0)$, was simply
rescaled as
\beq
\M^{\ZH,\si}_0 = \frac{s_{12}-\MZ^2}{s_{12}-\MZ^2+\ri\MZ\GZ} \; 
\M_0^{\ZH,\si}(\GZ=0).
\label{eq:MZH0GZ}
\eeq
This modification reproduces the correct Breit--Wigner shape near the
resonance and changes $\M_0^{\ZH,\si}(\GZ=0)$ at the relative order
${\cal O}(\GZ/\MZ)\sim {\cal O}(\alpha)$ away from the resonance.  It
should also be mentioned that we introduce a fixed Z-boson width (in
contrast to a running width), i.e.\ the Z-boson mass $\MZ$ deviates
from the on-shell mass at the two-loop level (see e.g.\ 
\citere{Bardin:1988xt}).

The issue of gauge invariance, resonances, Dyson summation, and
radiative corrections is rather complex.  In fact, no simple general
solution for a gauge-invariant treatment of finite-width effects
exists yet. Fortunately, the situation in our case is relatively
simple. We proceed in two different ways.

First we apply the so-called naive {\it fixed-width scheme} which
simply means that each resonance propagator $1/(s_{12}-\MZ^2)$ is
replaced by $1/(s_{12}-\MZ^2+\ri\MZ\GZ)$, while non-resonant
contributions are kept untouched. The contribution $\ri\MZ\GZ$
originates from the imaginary part of the (transverse part of the) Z
self-energy $\Sigma^{\PZ\PZ}_{\mathrm{T}}$ on resonance; in order to
avoid double-counting, the ${\cal O}(\alpha)$ contribution thus
contained in $\M^{\ZH,\si}_0$ has to be subtracted from the one-loop
amplitude.  In summary, the one-loop amplitude in the fixed-width
scheme reads
\beqar
\M_{1,\mbox{\scriptsize fixed-width}}^{\ZH,\si} &=&
\biggl[ \M_1^{\ZH,\si}(\GZ=0) - \M_{\mbox{\scriptsize ZZ-self}}^{\si}(\GZ=0) \biggr] 
_{\frac{1}{s_{12}-\MZ^2}\,\to\,\frac{1}{s_{12}-\MZ^2+\ri\MZ\GZ} } 
\nn\\[.5em]
&& {}
-\left[ \frac{\Sigma^{\PZ\PZ}_{\mathrm{T}}(s)
-\Re\{\Sigma^{\PZ\PZ}_{\mathrm{T}}(\MZ^2)\} }{s-\MZ^2}
+ \frac{\Sigma^{\PZ\PZ}_{\mathrm{T}}(s_{12})
-\Sigma^{\PZ\PZ}_{\mathrm{T}}(\MZ^2) }{s_{12}-\MZ^2}
+ 2\delta Z_{\PZ\PZ} \right] \M^{\ZH,\si}_0,
\nn\\
\eeqar
where we have made the Z self-energy and its corresponding counterterm
contributions explicit.  Note that the term
$\Re\{\Sigma^{\PZ\PZ}_{\mathrm{T}}(\MZ^2)\}$ is identical with the
mass counterterm $\delta\MZ^2$, while
$\Sigma^{\PZ\PZ}_{\mathrm{T}}(\MZ^2)$ receives its imaginary part
$\ri\Im\{\Sigma^{\PZ\PZ}_{\mathrm{T}}(\MZ^2)\}$ in the resonant part
from the subtraction of the $\ri\MZ\GZ$ contribution already contained
in $\M^{\ZH,\si}_0$.  Since only complete $S$-matrix elements exhibit
the gauge-invariance properties such as the independence of
gauge-fixing conditions (gauge parameters) and the validity of
Slavnov--Taylor identities, this procedure potentially violates gauge
invariance.

As a second option, we applied a {\it factorization scheme} where the
full $\PZ\PH$-production amplitude before Dyson summation is rescaled
similar to $\M_0^{\ZH,\si}(\GZ=0)$ in Eq.~\refeq{eq:MZH0GZ}.  This
procedure is analogous to the treatment of the W-boson resonance in
$\Pp\Pp\to\PW\to l\nu_l$ as described in \citere{Dittmaier:2001ay}.
Of course, also here double-counting of $\ri\MZ\GZ$ terms has to be
avoided.  In summary, the one-loop amplitude in the factorization
scheme reads
\beqar
\M_{1,\mathrm{fact.}}^{\ZH,\si} &=&
\frac{s_{12}-\MZ^2}{s_{12}-\MZ^2+\ri\MZ\GZ} 
\M_1^{\ZH,\si}(\GZ=0)
+ \frac{\ri\Im\{\Sigma^{\PZ\PZ}_{\mathrm{T}}(\MZ^2)\} }{s_{12}-\MZ^2}
 \M^{\ZH,\si}_0
\nn\\[.5em]
&=&
\frac{s_{12}-\MZ^2}{s_{12}-\MZ^2+\ri\MZ\GZ} \left[
\M_1^{\ZH,\si}(\GZ=0) - \M_{\mbox{\scriptsize ZZ-self}}^{\si}(\GZ=0) \right]
\nn\\[.5em]
&& {}
-\left[ \frac{\Sigma^{\PZ\PZ}_{\mathrm{T}}(s)
-\Re\{\Sigma^{\PZ\PZ}_{\mathrm{T}}(\MZ^2)\} }{s-\MZ^2}
+ \frac{\Sigma^{\PZ\PZ}_{\mathrm{T}}(s_{12})
-\Sigma^{\PZ\PZ}_{\mathrm{T}}(\MZ^2) }{s_{12}-\MZ^2}
+ 2\delta Z_{\PZ\PZ} \right] \M^{\ZH,\si}_0.
\nn\\
\eeqar
Note that this procedure preserves gauge invariance, since the
gauge-invariant amplitude $\M_1^{\ZH,\si}(\GZ=0)$ is only rescaled and
accompanied by another gauge-invariant term proportional to
$\M^{\ZH,\si}_0$.  However, the rescaling puts all non-resonant terms
in $\M_1^{\ZH,\si}(\GZ=0)$ to zero on the resonance. The corresponding
error is of the order of the non-resonant terms in the $\Oa$
corrections, \ie of order ${\cal O}(\alpha\GZ/\MZ)\sim {\cal
  O}(\alpha^2)$.

Within integration errors, both schemes give the same results.

\subsubsection{Universal electroweak corrections}
\label{se:univ_corr}

The electroweak corrections contain large contributions of universal
origin.  Besides initial-state radiation (ISR), which is discussed in
\refse{se:ISR}, these consist, in particular, of the corrections
associated with the running of $\al$ and corrections proportional to
$\Mt^2/\MW^2$ that can be associated to the $\rho$ parameter or to the
renormalization of the electroweak mixing angle. By suitable
parametrization of the lowest-order matrix elements, these universal
corrections can be incorporated into the lowest order thus reducing
the remaining corrections. This does not only reduce the $\Oa$
corrections but in general also the higher-order corrections.

The running of $\alpha(Q^2)$ from $Q^2=0$ to $Q^2=\MZ^2$ is of the
order of $\Delta\alpha\sim6\%$.  Since the cross section for $\eennh$
is proportional to $\alpha^3$ this amounts to an effect of $\sim18\%$
if $\alpha(0)$ defines the electromagnetic coupling
(``$\alpha(0)$-scheme'').  By parametrizing the lowest-order cross
section in terms of $\alpha(\MZ^2)$ (``$\alpha(\MZ^2)$-scheme'') this
large effect can be incorporated into the leading-order expressions.
Note that $\Delta\alpha$ contains large contributions from the
hadronic vacuum polarization that cannot be calculated in perturbation
theory.  Thus, the sensitivity to these effects is absorbed in the
lowest-order cross section via $\alpha(\MZ^2)$ in the
$\alpha(\MZ^2)$-scheme.  Alternatively the electromagnetic coupling
can be deduced from the Fermi constant $\GF$ (``$\GF$-scheme'') via
\beq\label{eq:defGF}
\alpha_{\GF} = \frac{\sqrt{2}\GF\MW^2\sw^2}{\pi},
\eeq
where $\GF$ is measured in muon decay. Consequently, in the transition
from the $\alpha(0)$- to the $\GF$-scheme the constant $3\Delta r$ has
to be subtracted from the relative correction to a cross section that
is proportional to $\alpha^3$, where $\Delta r$ contains the
electroweak radiative corrections to muon decay.  Since $\Delta r$ is
about 3\% for the empirical value of $\Mt$, this shifts the relative
corrections by $\sim9\%$ with respect to the $\alpha(0)$-scheme. Apart
from absorbing $\Delta\alpha$, the quantity $\Delta r$ additionally
contains universal corrections proportional to $\Mt^2/\MW^2$ that can
be associated to the $\rho$ parameter, $\Delta
r-\Delta\alpha\sim-\Delta\rho\,\cw^2/\sw^2\sim -3\%$ with
$\Delta\rho=3\GF\Mt^2/(8\sqrt{2}\pi^2)$ \cite{Burgers:1989bh}.

The cross section for $\eennh$ is dominated by the $\PW\PW$-fusion
diagram, which gets its main contribution from the region of small
momentum transfers. Consequently, the corresponding corrections are
determined by the $\Pe\Pne\PW$ and $\PW\PW\PH$ vertex corrections for
small invariant $\PW$ masses and depend only weakly on the energy.
The correction to the $\Pe\Pne\PW$ vertex in the relevant kinematical
region is well approximated by $\Delta r$. It turns out that this is
also the case for the main contributions to the $\PW\PW\PH$ vertex.
Thus, parametrizing the lowest order in terms of $\GF$ ($\GF$-scheme)
absorbs a large part of the universal corrections.  Since in the
$\GF$-scheme all large universal corrections related to the running of
$\al$ and most of the corrections $\propto\Mt^2/\MW^2$ are absorbed,
we use this scheme in the following. With respect to this scheme, the
corrections in the $\alpha(0)$-scheme are shifted by $3\Delta r\approx
+9\%$ and those in the $\alpha(\MZ^2)$-scheme by $3(\Delta
r-\Delta\alpha(\MZ^2))\approx -9\%$.

We have extracted the leading $\Mt$-dependent corrections in the
heavy-top limit in the  $\GF$-scheme:
\beqar
\M_1^{\ZH,\si}\Big|_{\GF\mbox{\scriptsize-scheme}} &\;\asymp{\Mt\to\infty}\; &
\frac{\al}{4\pi\sw^2}
\left[
\frac{1}{8}\left(\frac{6\cw}{\sw}+g^\si_{\PZ\Pe}\right) \, \frac{\Mt^2}{\MW^2} 
+\frac{3-2\sw^2}{3\cw\sw} \ln\frac{\Mt}{\MW}
\right] \frac{\M_0^{\ZH,\si}}{g^\si_{\PZ\Pe}} 
\nn\\[1ex]&& {} 
\;+\; {\cal O}(\Mt^0),
\nn\\[.5em]
\M_1^{\WW}\Big|_{\GF\mbox{\scriptsize-scheme}} &\;\asymp{\Mt\to\infty} \;&
-\frac{5\al}{32\pi\sw^2} \, \frac{\Mt^2}{\MW^2} \, \M_0^{\WW} 
\;+\; {\cal O}(\Mt^0).
\eeqar
The leading $\Mt^2$-enhanced terms of the WW channel agree with the
terms derived for the $\PH\PW\PW$ vertex \cite{Kniehl:1995at}, since
in the $\GF$-scheme all leading $\Mt^2$ contributions related to the
W-boson coupling to fermions are absorbed in $\al_{\GF}$.  In the ZH
channel, $\Mt^2$-enhanced terms do not only result from the
$\PH\PZ\PZ$ vertex, but there are also remnants originating from the
renormalization of the Z-boson couplings to fermions. In contrast to
the WW channel, in the ZH channel there are also logarithmic terms
$\ln\Mt$ for a large top-quark mass.

In the physically interesting region of $\eennh$, $\sqrt{s}$ is of the
same order as $\Mt$ or larger. Nevertheless, the expressions for
$\Mt\to\infty$ reproduce the full $\Mt$-dependent corrections rather
well for the WW channel, which is dominated by small momentum
transfers. The situation for the ZH channel, where $\sqrt{s}$ is a
typical scale for the momentum transfer, is completely different.
There, the expressions for $\Mt\to\infty$ do not provide a good
approximation, and we did not succeed in finding a simple
approximation for the fermionic corrections to the ZH channel. This is
due to the presence of loop integrals that depend both on $s$ and
$\Mt^2$. Since, however, the cross section for $\eennh$ is dominated
by the WW channel we consider it useful to define the following
improved Born approximation
\beq
\rd\sigma_{\IBA}^{\mbox{\scriptsize non-photonic}}=
\rd\sigma_{0}-\rd\sigma^{\WW}_{0}\frac{5\al}{16\pi\sw^2} \, \frac{\Mt^2}{\MW^2}.
\label{eq:IBAnon-phot}
\eeq
This cross section is then convoluted with the ISR as given in 
\refeq{sigmaLLIBA} below.

As discussed in \citere{Denner:2003yg} and in \refse{se:numres}, the
bosonic corrections are small for the WW channel but large for the ZH
channel.  In order to find the source of these large corrections, we
have evaluated the leading bosonic corrections in the limit
$s\sim\MH^2\gg\MW^2$ which are of the order ${\cal
  O}(s/\MW^2)\sim{\cal O}(\MH^2/\MW^2)$.  We found that these terms
are small and cannot explain the large bosonic corrections in the ZH
channel. In the 't~Hooft--Feynman gauge the large corrections arise
predominantly from box diagrams involving W-boson exchange
\cite{Denner:1992bc}.  The enhancement of these diagrams is partially
due to the gauge-boson coupling to the electron which for W~bosons is
stronger than for Z~bosons.

\subsection{Real photonic corrections}
\label{se:rrcs}

\subsubsection{Matrix-element calculation}
\label{se:bremsme}

The real photonic corrections are induced by the process
\beq
\Pem(p_1,\sigma_1) + \Pep(p_2,\sigma_2) \;\longrightarrow\;
\nu_l(k_1) + \bar\nu_l(k_2) + \PH(k_3) + \gamma(k,\lambda), 
\qquad l=\Pe,\mu,\tau,
\label{eq:eennha}
\eeq
where $k$ and $\lambda$ denote the photon momentum and helicity,
respectively. The Feynman diagrams of this process are shown
in \reffi{eennHgamma}.
\begin{figure}
\centerline{\footnotesize  \input{paper-real}}
\caption{Feynman diagrams for $\Pep\Pem\to\nu\bar\nu\PH\ga$}
\label{eennHgamma}
\end{figure}
We have evaluated the helicity matrix elements $\M^{\si\la}_\ga$ of
this process using the Weyl--van der Waerden spinor technique as
formulated in \citere{Dittmaier:1999nn}.  The amplitudes for the
helicity channels with $\si_1=\si_2$ vanish for massless electrons.
The finite-mass effects of electrons and positrons are included in the
treatment of soft and collinear singularities, as described in the
next section.  In the notation of \citere{Dittmaier:1999nn} the
amplitudes $\M^{\si\la}_\ga$ read
\beqar\label{eq:MEnnHg}
\M^{\si\la}_\ga &=& \M^{\ZH,\si\la}_{\ga} 
+ \delta_{\si-} \, \M^{\WW,\la}_{\ga},
\nn\\[.5em]
\M^{\ZH,\si\la}_{\ga} &=& \frac{\sqrt{2}e^4g^\si_{\PZ\Pe}\MW}{\cw^3\sw^2} \,
\frac{1}{(p_1+p_2-k)^2-\MZ^2} \,
\frac{1}{(k_1+k_2)^2-\MZ^2+\ri\MZ\GZ} \,
\nn\\
&& {} \times A_{\si\la}^{\ZH}(p_1,p_2,k_1,k_2,k),
\nn\\[.5em]
\M^{\WW,\la}_{\ga} &=&
\delta_{l\Pe} \, \frac{\sqrt{2}e^4\MW}{\sw^3} \,
\frac{1}{(p_1-k_1)^2-\MW^2} \,
\frac{1}{(p_2-k_2)^2-\MW^2} \,
\nn\\
&& {} \times 
A_{\la}^{\WW}(p_1,p_2,k_1,k_2,k)
\eeqar
with the auxiliary functions
\beqar\label{eq:MEnnHgaux}
A_{--}^{\ZH}(p_1,p_2,k_1,k_2,k) &=& -\langle p_1 k_2\rangle^*
\frac{ \langle p_1 p_2 \rangle^* \langle p_2 k_1\rangle
      +\langle p_1 k   \rangle^* \langle k_1 k  \rangle }
     {\langle p_1 k \rangle^* \langle p_2 k \rangle^* },
\nn\\
A_{\si,-\la}^{\ZH}(p_1,p_2,k_1,k_2,k) &=& 
-A_{\si\la}^{\ZH}(p_2,p_1,k_2,k_1,k)^*, 
\nn\\
A_{-\si,-\la}^{\ZH}(p_1,p_2,k_1,k_2,k) &=& 
\phantom{-}A_{\si\la}^{\ZH}(p_1,p_2,k_2,k_1,k)^*, 
\nn\\[.5em]
A_{-}^{\WW}(p_1,p_2,k_1,k_2,k) &=&
\langle p_1 k_2\rangle^* \Biggl\{
-\frac{ \langle p_1 p_2 \rangle^*\langle p_2 k_1\rangle}
      {\langle p_1 k \rangle^* \langle p_2 k \rangle^* }
+\frac{ \langle k_1 k  \rangle 
        \left(\langle p_1 k_1 \rangle^* \langle p_2 k_1\rangle
       +\langle p_1 k \rangle^* \langle p_2 k \rangle\right)}
      { \langle p_1 k \rangle^*\left[(p_1-k_1-k)^2-\MW^2\right]}
\nn\\
&& \phantom{ \langle p_1 k_2\rangle^* \Biggl\{ } {}
-\frac{ \langle k_1 k\rangle} {\langle p_2 k \rangle^* }
-\frac{ \langle k_2 k  \rangle 
        \left(\langle p_2 k_2 \rangle^* \langle p_2 k_1\rangle
       -\langle k_2 k \rangle^* \langle k_1 k \rangle\right)}
      { \langle p_2 k \rangle^*\left[(p_2-k_2-k)^2-\MW^2\right]} \Biggr\},
\nn\\
A_{-\la}^{\WW}(p_1,p_2,k_1,k_2,k) &=&
-A_{\la}^{\WW}(p_2,p_1,k_2,k_1,k)^*.
\eeqar
The relations between the $A^{\dots}_{\dots}$ functions that differ
only in the photon helicity $\la$ are a consequence of the CP symmetry
of the process, while the relation associated with a reversion of both
$\si$ and $\la$ results from a P transformation.  The spinor products
$\langle\dots\rangle$ are defined by
\beq
\langle pq\rangle=\epsilon^{AB}p_A q_B
=2\sqrt{p_0 q_0} \,\Biggl[
{\mathrm{e}}^{-\ri\phi_p}\cos\frac{\theta_p}{2}\sin\frac{\theta_q}{2}
-{\mathrm{e}}^{-\ri\phi_q}\cos\frac{\theta_q}{2}\sin\frac{\theta_p}{2}
\Biggr],
\eeq
where $p_A$, $q_A$ are the associated momentum spinors for the light-like
momenta
\beqar
p^\mu&=&p_0(1,\sin\theta_p\cos\phi_p,\sin\theta_p\sin\phi_p,\cos\theta_p),\nl
q^\mu&=&q_0(1,\sin\theta_q\cos\phi_q,\sin\theta_q\sin\phi_q,\cos\theta_q).
\eeqar

The matrix elements \refeq{eq:MEnnHg}, \refeq{eq:MEnnHgaux} have been
successfully checked against the result obtained with the package {\sl
  Madgraph} \cite{Stelzer:1994ta} numerically.

The contribution $\sigma_\gamma$ of the radiative process to the cross
section is given by
\beq
\sigma_\gamma = \frac{1}{2s} \int \rd\Phi_\gamma \,
 \sum_{\si=\pm\frac{1}{2}} \frac{1}{4}
(1+2P_-\si)(1-2P_+\si) \, \sum_{\lambda=\pm 1} \,
|\M^{\si\la}_\gamma|^2,
\label{eq:hbcs}
\eeq
where the phase-space integral is defined by
\beq
\int \rd\Phi_\gamma =
\int\frac{\rd^3 {\bf k}}{(2\pi)^3 2k^0} \,
\left( \prod_{i=1}^3 \int\frac{\rd^3 {\bf k}_i}{(2\pi)^3 2k_i^0} \right)\,
(2\pi)^4 \delta\Biggl(p_1+p_2-k-\sum_{j=1}^3 k_j\Biggr).
\label{eq:dPSg}
\eeq

\subsubsection{Treatment of soft and collinear singularities}

Without soft and collinear regulators the phase-space integral
\refeq{eq:hbcs} diverges in the soft ($k_0\to 0$) and collinear ($p_i
k\to 0$) phase-space regions.  In the following we describe two
procedures of treating soft and collinear photon emission: one is
based on a subtraction method, the other on phase-space slicing. In
both cases soft and collinear singularities are regularized by an
infinitesimal photon mass and a small electron mass, respectively.

\paragraph{The dipole subtraction approach}

The idea of so-called subtraction methods is to subtract a simple
auxiliary function from the singular integrand of the bremsstrahlung
integral and to add this contribution back again after partial
analytic integration.  This auxiliary function, denoted $|\M_\sub|^2$
in the following, has to be chosen in such a way that it cancels all
singularities of the original integrand, which is $|\M_\ga|^2$ in our
case, so that the phase-space integration of the difference can be
performed numerically, even over the singular regions of the original
integrand.  In this difference, $\M_\ga$ can be evaluated without
regulators for soft or collinear singularities, i.e.\ we can make use
of the results of the previous section.  The auxiliary function has to
be simple enough so that it can be integrated over the singular
regions analytically, when the subtracted contribution is added again.
This part contains the singular contributions and requires regulators,
i.e.\ photon and electron masses have to be reintroduced there.
Specifically, we have applied the {\it dipole subtraction formalism},
which is a process-independent approach that was first proposed
\cite{Catani:1996jh} within QCD for massless unpolarized partons and
subsequently generalized to photon radiation of massive polarized
fermions \cite{Dittmaier:2000mb}.  We only need the limit of small
fermion masses, which was worked out in
\citeres{Dittmaier:2000mb,Roth:1999kk} and in which the application of
the method is relatively simple.  In order to keep the description of
the method transparent, we describe only the basic structure of the
individual terms explicitly and refer to
\citeres{Dittmaier:2000mb,Roth:1999kk} for details.

In the dipole subtraction formalism the subtraction function is
constructed from contributions that are labelled by ordered pairs $ab$
of charged fermions, so-called ``dipoles''. The fermions $a$ and $b$
are called {\it emitter} and {\it spectator}, respectively, since by
construction only the kinematics of the emitter $a$ leads to collinear
singularities.  Since we only have charged particles in the initial
state, the subtraction function receives two contributions, which both
have emitter and spectator in the initial state,
\beqar
|\M^\si_\sub(p_1,p_2,k_i,k)|^2 &=& \sum_{a,b=1,2 \atop a\ne b}
|\M^\si_{\sub,ab}(p_1,p_2,k_i,k)|^2 ,
\nn\\
|\M^\si_{\sub,ab}(p_1,p_2,k_i,k)|^2 &=&
e^2 \gsub_{ab}(p_a,p_b,k)
\left|\M^\si_0(\tilde p_1,\tilde p_2,\tilde k_i)\right|^2
\label{eq:msub}
\eeqar
with the dipole function
\beq
\gsub_{ab}(p_a,p_b,k) =
\frac{1}{(p_a k)x_{ab}} \biggl[ \frac{2}{1-x_{ab}}-1-x_{ab} \biggr],
\eeq
where
\beq
x_{ab} = \frac{p_a p_b-p_a k-p_b k}{p_a p_b}.
\eeq
The modified momenta in Eq.~\refeq{eq:msub} 
depend on $ab$ and are defined as follows.
While the spectator momentum $p_b$ is kept fixed,
the emitter momentum $p_a$, is rescaled by $x_{ab}$,
\beq
\tilde p_a^\mu = x_{ab} p_a^\mu, \qquad 
\tilde p_b^\mu = p_b^\mu,
\eeq
and all other momenta $k_i$ are transformed with a Lorentz transformation,
\beq
\tilde k_j^{\mu} = \Lambda^{\mu}_{\phantom{\mu}\nu} k_j^\nu,
\label{eq:tkj}
\eeq
where
\beqar
\Lambda^{\mu}_{\phantom{\mu}\nu} &=& g^\mu_{\phantom{\mu}\nu}
-\frac{(P_{ab}+\tilde P_{ab})^\mu(P_{ab}+\tilde P_{ab})_\nu}
{P_{ab}^2+P_{ab}\tilde P_{ab}}
+\frac{2\tilde P_{ab}^\mu P_{ab,\nu}}{P_{ab}^2},
\nn\\
P_{ab} &=& p_a + p_b - k = \sum_j k_j, \qquad
\tilde P_{ab}^\mu = x_{ab} p_a^\mu + p_b^\mu.
\label{eq:LT}
\eeqar

The subtracted contribution can be integrated over the (singular)
photonic degrees of freedom up to a remaining convolution over
$x=x_{ab}$.  In this integration the regulators $m_\gamma$ and $\Me$
must be retained, and the soft and collinear singularities appear as
logarithms in these mass regulators,%
\beqar
\sigma_{\sub}(p_1,p_2,P_-,P_+) &=&
\frac{\alpha}{2\pi} \int_0^1\rd x\,
\sum_{\tau=\pm} \, \cGsub_{\tau}(s,x)\Biggl[ \phantom{{}+{}}
\int\rd\sigma_0(xp_1,p_2,\tau P_-,P_+) 
\nn\\
&& \phantom{\frac{\alpha}{2\pi} \int_0^1\rd x\,
\sum_{\tau=\pm} \, \cGsub_{\tau}(s,x)\Bigl[}
+ \int\rd\sigma_0(p_1,xp_2,P_-,\tau P_+) \Biggr]
\nn\\ && {}
+ \frac{\alpha}{2\pi} \sum_{\tau=\pm} \,
\Gsub_\tau(s) \Biggl[ 
\int\rd\sigma_0(s,\tau P_-,P_+) + \int\rd\sigma_0(s,P_-,\tau P_+) \Biggr],
\nn\\
\eeqar
with the universal functions
\beqar
\cGsub_+(s,x) &=&
\left(\frac{1+x^2}{1-x}\right)_+
\left[\ln\biggl(\frac{s}{\Me^2}\biggr)-1\right], \qquad
\cGsub_-(s,x) = (1-x)_+,
\nn\\
\Gsub_+(s) &=& 
\ln\biggl(\frac{\Me^2}{s}\biggr)
\ln\biggl(\frac{m_\gamma^2}{s}\biggr)
+ \ln\biggl(\frac{m_\gamma^2}{s}\biggr)
- \frac{1}{2}\ln^2\biggl(\frac{\Me^2}{s}\biggr)
+ \frac{1}{2}\ln\biggl(\frac{\Me^2}{s}\biggr)
-\frac{\pi^2}{3} + \frac{3}{2},
\nn\\
\Gsub_-(s) &=& \frac{1}{2}.
\eeqar
The two cases $\tau=+/-$ correspond to collinear photon emission
without/with a spin-flip of the electron or positron.
The $(\dots)_+$ prescription is defined as usual,
\beq
\int_0^1\rd x\, \Big(f(x)\Big)_+ g(x) \equiv
\int_0^1\rd x\, f(x) \left[g(x)-g(1)\right],
\eeq
and $P_\pm$ are the degrees of polarization of the $\Pe^\pm$ beams.

In summary, the phase-space integral \refeq{eq:hbcs} in the dipole
subtraction approach reads
\beq
\sigma_\gamma = \frac{1}{2s} \int \rd\Phi_\gamma \,
\sum_{\si=\pm\frac{1}{2}} \frac{1}{4} 
(1+2P_-\si)(1-2P_+\si) \, 
\left[ \sum_{\lambda=\pm 1} \, |\M^{\si\la}_\gamma|^2
-|\M^\si_\sub|^2 \right] \;+\; \sigma_{\sub}.
\eeq

\paragraph{The phase-space-slicing approach}

The idea of the phase-space-slicing method is to divide the
bremsstrahlung phase space into singular and non-singular regions,
then to evaluate the singular regions analytically and to perform an
explicit cancellation of the arising soft and collinear singularities
against their counterparts in the virtual corrections. The finite
remainder can be evaluated by using the usual Monte Carlo techniques.
For the actual implementation of this well-known procedure we closely
follow the approaches of \citere{bo93}.  We divide the five-particle
phase space into soft and collinear regions by introducing the cut-off
parameters $\De E$ and $\De\theta$, respectively. We decompose the
real corrections as
\begin{equation}
\rd\sigma_{\ga} = \rd\sigma_{\soft}+ \rd\sigma_{\coll}+\rd\sigma_{\ga,\finite}.
\end{equation}
Here $\rd\sigma_{\soft}$ describes the contribution of the soft
photons, \ie of photons with energies $k_0 < \Delta E$ in the CM
frame, and $\rd\sigma_{\coll}$ describes real photon radiation outside
the soft-photon region ($k_0>\Delta E$) but collinear to the $\Pe^\pm$
beams.  The collinear region consists of the two disjoint parts
$0<\theta_{\ga}<\De\theta$ and $\pi-\De\theta<\theta_{\ga}<\pi$, where
$\theta_{\ga}$ is the polar angle of the emitted photon in the CM
frame. The remaining part, which is free of singularities, is denoted
by $\rd\sigma_{\ga,\finite}$.

In the soft and collinear regions, the squared matrix element
$|\M_{\ga}|^2$ factorizes into the leading-order squared matrix
element $|\M_0|^2$ and a soft or collinear factor.  Also the
four-particle phase space factorizes into a three-particle and a soft
or collinear part, so that the integration over the singular part of
the photon phase space can be performed analytically.

In the soft-photon region, we apply the soft-photon approximation to
$|\M_{\ga}|^2$, \ie the photon four-momentum $k$ is omitted
everywhere but in the IR-singular propagators. In this region
$\rd\sigma_{\ga}$ can be written as \cite{Denner:1993kt,Yennie:1961ad}
\beq
\rd\sigma_{\soft} = - \rd\sigma_0
\frac{\alpha}{4\pi^2} 
\int_{k_0< \Delta E \atop |{\bf k}|^2=k_0^2-m_\ga^2}
\frac{\rd^3 {\bf k}}{k_0}
\left(\frac{p_1^{\mu}}{p_1 k}-\frac{p_2^{\mu}}{p_2 k}\right)^2.
\eeq
The explicit expression for the soft-photon integral can be found in
\citeres{'tHooft:1979xw,Denner:1993kt}.  For our purpose it is
sufficient to keep the electron mass only as regulator for the
collinear singularities. In this limit we obtain
\beq
\rd\sigma_{\soft} = - \rd\sigma_0 \, \frac{\alpha}{\pi}
\left\{ 2 \ln\left(\frac{2\Delta E}{m_\ga}\right)
\, \left[1-\ln\left(\frac{s}{\Me^2}\right)\right]
-\ln\left(\frac{s}{\Me^2}\right)
+\frac{1}{2}\ln^2\left(\frac{s}{\Me^2}\right)+
\frac{\pi^2}{3}
\right\}.
\label{eq:si_soft}
\eeq

In the collinear region, we consider an incoming $\Pe^\mp$ with
momentum $p_i$ being split into a collinear photon and an $\Pe^\mp$
with the resulting momentum $xp_i$ after photon radiation.  In the
asymptotic limit, $|\M_{\ga}|^2$ factorizes into the leading-order
squared matrix element $|\M_0|^2$ and a collinear factor describing
collinear initial-state radiation, as long as $\De\theta$ is
sufficiently small.  In the collinear region also the four-particle
phase space factorizes into a three-particle phase space and a
collinear part, so that the cross section for hard photon radiation
($k_0>\Delta E$) in the collinear region reads
\beqar
\sigma_{\coll}(p_1,p_2,P_-,P_+) &=&
\frac{\alpha}{2\pi} \int_0^{1-2\De E/\sqrt{s}}\rd x\,
\sum_{\tau=\pm} \, \cGcoll_{\tau}(s,x)\Biggl[ \phantom{{}+{}}
\int\rd\sigma_0(xp_1,p_2,\tau P_-,P_+) 
\nn\\*
&& \phantom{\frac{\alpha}{2\pi} \int_0^{1-2\De E/\sqrt{s}}\rd x\,
\sum_{\tau=\pm} \, \cGcoll_{\tau}(s,x)\Bigl[}
+ \int\rd\sigma_0(p_1,xp_2,P_-,\tau P_+) \Biggr],
\nn\\
\label{eq:si_collin}
\eeqar
where
\beqar
\cGcoll_+(s,x) &=&
\frac{1+x^2}{1-x}
\left[\ln\biggl(\frac{s\De\theta^2}{4\Me^2}\biggr)-1\right], \qquad
\cGcoll_-(s,x) = 1-x.
\eeqar

Subtracting the soft and collinear cross sections \refeq{eq:si_soft}
and \refeq{eq:si_collin} from the cross section of the bremsstrahlung
process yields the finite cross section $\rd\sigma_{\ga,\finite}$.  As
usual in the phase-space-slicing approach, this subtraction is done in
practice by imposing cuts on the bremsstrahlung phase space, \ie a
photon-energy cut, $k_0>\Delta E$, and a cut on the angles between the
photon and the beams, $\De\theta<\theta_{\ga}<\pi-\De\theta$.

\subsubsection{Initial-state radiation in ${\cal O}(\alpha)$ and beyond}
\label{se:ISR}

In $\Oa$ the effect of ISR is entirely contained in the radiative
corrections described above. However, the emission of photons
collinear to the incoming electrons or positrons leads to corrections
that are enhanced by large logarithms.  In order to achieve an
accuracy at the few $0.1\%$ level, the corresponding higher-order
contributions, i.e.\ contributions beyond $\Oa$, must be taken into
account. This can be done in the structure-function method
\cite{sf,lep2repWcs}.  According to the mass-factorization theorem,
the leading-logarithmic (LL) initial-state QED corrections can be
written as a convolution of the lowest-order cross section with
structure functions, and the corresponding differential cross section
reads
\newcommand{\LL}{\mathrm{LL}}
\newcommand{\ISR}{\mathrm{ISR}}
\beq\label{sigmaLL}
  \int \rd\sigma_{\ISR,\LL} =
  \int^1_0 \rd x_1 \int^1_0 \rd x_2 \,
  \Gamma_{\Pe\Pe}^{\LL}(x_1,Q^2)\Gamma_{\Pe\Pe}^{\LL}(x_2,Q^2)
  \int \rd\sigma_0(x_1 p_1,x_2 p_2).
\eeq
Here $x_1$ and $x_2$ denote the fractions of the longitudinal momentum
carried by the incoming electron and positron momenta just before the
hard scattering process occurs.  This means that the incoming momenta
$p_\pm$ before emission of the collinear photon are rescaled by
$x_{1,2}$, and the CM frame of the hard scattering process with the
lowest-order cross section $\rd\sigma_0(x_1 p_1,x_2 p_2)$ is boosted
along the beam axis.  The LL structure function including $\Oaaa$
terms is given by \cite{lep2repWcs}
\newcommand{\zetal}{\be_\Pe}
\beqar\label{SFexp}
  \Gamma_{\mathrm{ee}}^{\LL}(x,Q^2) &=&
    \frac{\exp\left(-\frac{1}{2}\zetal\gamma_{\rE} +
        \frac{3}{8}\zetal\right)}
{\Gamma\left(1+\frac{1}{2}\zetal\right)}
    \frac{\zetal}{2} (1-x)^{\frac{\zetal}{2}-1} - \frac{\zetal}{4}(1+x)
\nn\\
&&  {} - \frac{\zetal^2}{32} \biggl\{ \frac{1+3x^2}{1-x}\ln(x)
    + 4(1+x)\ln(1-x) + 5 + x \biggr\}
\nn\\
&&  {} - \frac{\zetal^3}{384}\biggl\{
      (1+x)\left[6\Li(x)+12\ln^2(1-x)-3\pi^2\right]
\nn\\
&& \quad\quad {}
+\frac{1}{1-x}\biggl[ \frac{3}{2}(1+8x+3x^2)\ln(x)
+6(x+5)(1-x)\ln(1-x)
\nn\\
&& \quad\quad\quad {}
+12(1+x^2)\ln(x)\ln(1-x)-\frac{1}{2}(1+7x^2)\ln^2(x)
\nn\\
&& \quad\quad\quad  {}
+\frac{1}{4}(39-24x-15x^2)\biggr] \biggr\}
\eeqar
with
\beq
  \zetal = \frac{2\alpha}{\pi} \left(L-1\right)
\eeq
and the leading logarithm
\beq
L = \ln\frac{Q^2}{\Me^2}.
\eeq
Note that the scale $Q^2$ is not fixed within LL approximation, but
has to be set to a typical scale of the underlying process; for the
numerics we use $Q^2=s$.  In \refeq{SFexp} $\gamma_{\rE}$ is the Euler
constant and $\Gamma(y)$ the gamma function, which should not be
confused with the structure functions.  Note that some non-leading
terms are incorporated, taking into account the fact that the residue
of the soft-photon pole is proportional to $L-1$ rather than $L$ for
the initial-state photon radiation.

We add the cross section \refeq{sigmaLL} to the one-loop result and
subtract the lowest-order and one-loop contributions
$\rd\sigma_{\ISR,\LL,1}$ already contained within this formula,
\beqar\label{sigmaLL1}
  \int \rd\sigma_{\ISR,\LL,1} &=&
  \int^1_0 \rd x_1\rd x_2
  \Bigl[\de(1-x_1)\de(1-x_2)
  +\Gamma_{\Pe\Pe}^{\LL,1}(x_1,Q^2)\de(1-x_2)
\nl
  &&\qquad
  {}+\de(1-x_1)\Gamma_{\Pe\Pe}^{\LL,1}(x_2,Q^2)\Bigr]
  \int \rd\sigma_0(x_1p_1,x_2p_2),
\eeqar
in order to avoid double counting.  The one-loop contribution to the
structure function reads
\beqar
  \Gamma_{\mathrm{ee}}^{\LL,1}(x,Q^2) &=&
  \frac{\zetal}{4} \left(\frac{1+x^2}{1-x}\right)_+
\nl
  &=& \frac{\zetal}{4} \lim_{\eps\to 0+}
  \left[\delta(1-x)\left(\frac{3}{2}+2\ln\eps\right)
  + \theta(1-x-\eps)\frac{1+x^2}{1-x}\right].
\eeqar
Note that the uncertainty that is connected with the choice of $Q^2$
enters only in ${\cal O}(\alpha^2)$, if all $\Oa$ corrections,
including constant terms, are taken into account.

Finally, we complete the definition of the IBA by dressing
$\rd\sigma_{\IBA}^{\mbox{\scriptsize non-photonic}}$ of
\refeq{eq:IBAnon-phot} with the ISR structure functions,
\beq\label{sigmaLLIBA}
  \int \rd\sigma_{\IBA} =
  \int^1_0 \rd x_1 \int^1_0 \rd x_2 \,
  \Gamma_{\Pe\Pe}^{\LL}(x_1,Q^2)\Gamma_{\Pe\Pe}^{\LL}(x_2,Q^2)
  \int \rd\sigma_{\IBA}^{\mbox{\scriptsize non-photonic}}(x_1 p_1,x_2 p_2).
\label{eq:IBA}
\eeq

\subsubsection{Monte Carlo integration}

The phase-space integration is performed with Monte Carlo techniques
in both computer codes. The first code employs a multi-channel Monte
Carlo generator similar to the one implemented in {\sl RacoonWW}
\cite{Roth:1999kk,Denner:1999gp} and {\sl Lusifer}
\cite{Dittmaier:2002ap}, the second one uses the adaptive
multi-dimensional integration program {\sl VEGAS}
\cite{Lepage:1977sw}.

\section{Numerical results}
\label{se:numres}

\subsection{Input parameters}

For the numerical evaluation we use the following set of SM parameters
\cite{Hagiwara:pw},
\beq
\begin{array}[b]{lcllcllcl}
\GF & = & 1.16639 \times 10^{-5} \GeV^{-2}, \ \ &
\alpha(0) &=& 1/137.03599976, \\
\MW & = & 80.423\GeV, &
\MZ^{\LEP} & = & 91.1876\GeV, &
\GZ^{\LEP} & = & 2.4952\GeV, \\
\Me & = & 0.510998902\MeV, &
m_\mu &=& 105.658357\MeV,\ \  &
m_\tau &=& 1.77699\GeV, \\
\Mu & = & 66\MeV, &
\Mc & = & 1.2\GeV, &
\Mt & = & 174.3\;\GeV, \\
\Md & = & 66\MeV, &
\Ms & = & 150\MeV, &
\Mb & = & 4.3\GeV,
\end{array}
\label{eq:SMpar}
\eeq
which coincides with the one used in \citere{Denner:2003yg}.  Since we
employ the on-shell renormalization scheme, the weak mixing angle is
fixed by \refeq{eq:defcw}.

As discussed in \refse{se:univ_corr}, and if not stated otherwise, we
evaluate amplitudes in the so-called $\GF$-scheme, i.e.\ we derive the
electromagnetic coupling $\alpha=e^2/(4\pi)$ from the Fermi constant
$\GF$ according to \refeq{eq:defGF}.  This procedure, in particular,
absorbs all sizeable mass effects of light fermions other than
electrons in the coupling $\alpha_{\GF}$, and the results are
practically independent of the masses of the light quarks. The masses
of the light quarks are adjusted to reproduce the hadronic
contribution to the photonic vacuum polarization of
\citere{Jegerlehner:2001ca}.  In the relative radiative corrections,
we use $\alpha(0)$ as coupling parameter, which is the correct
effective coupling for real photon emission. We do not calculate the
W-boson mass from $\GF$ but use its experimental value as input.

As explained in \refse{se:resonance}, we employ a fixed width in the
resonant Z-boson propagator in contrast to the approach used at LEP to
fit the Z~resonance, where a running width is taken.  Therefore, we
have to convert the ``on-shell'' values of $\MZ^{\LEP}$ and
$\GZ^{\LEP}$, resulting from LEP, to the ``pole values'' denoted by
$\MZ$ and $\GZ$ in this paper. The relation of the two sets of values
is given by \cite{Bardin:1988xt}
\beqar
\MZ &=& \MZ^{\LEP}/
\sqrt{1+(\GZ^{\LEP}/\MZ^{\LEP})^2} = 91.1535\GeV,
\nn\\
\GZ &=& \GZ^{\LEP}/
\sqrt{1+(\GZ^{\LEP}/\MZ^{\LEP})^2} = 2.4943\GeV,
\label{eq:zparam}
\eeqar
i.e.\ the difference is formally of 
two-loop order and numerically
hardly visible in the results presented below.

\subsection{Results on total cross sections}

In \citere{Denner:2003yg} we have already discussed numerical results
for the total cross sections of the process $\eennh$ and the
corresponding radiative corrections relative to the pure lowest-order
prediction. Particular attention has been paid to the individual
contributions of the various sources of corrections, such as the
contributions of the closed fermion loops, of the ISR effects in $\Oa$
and beyond, and of the remaining bosonic corrections. In this
discussion the ZH-production and WW-fusion channels have been
considered separately, thereby revealing characteristic differences of
the two channels.  In the following we continue the discussion of
\citere{Denner:2003yg}.  We always sum over all three neutrino
species, \ie over the processes $\eeneneh$, $\nu_\mu\bar\nu_\mu H$,
and $\nu_\tau\bar\nu_\tau H$.  Besides the full cross section, denoted
``total'' in the plots, in some cases we also give the cross section
resulting from the $\PZ\PH$-production and $\PW\PW$-fusion channels
separately, which are referred to as ``ZH'' and ``WW'' contributions,
respectively. In the $\PZ\PH$-production channel we sum over the
relevant contributions of all $\nu\bar\nu\PH$ final states, which is
equivalent to multiplying the cross sections for $\eenmnmh$ by a
factor 3.

\begin{figure}
\centerline{
\includegraphics[bb=40 415 250 625,width=.5\textwidth]{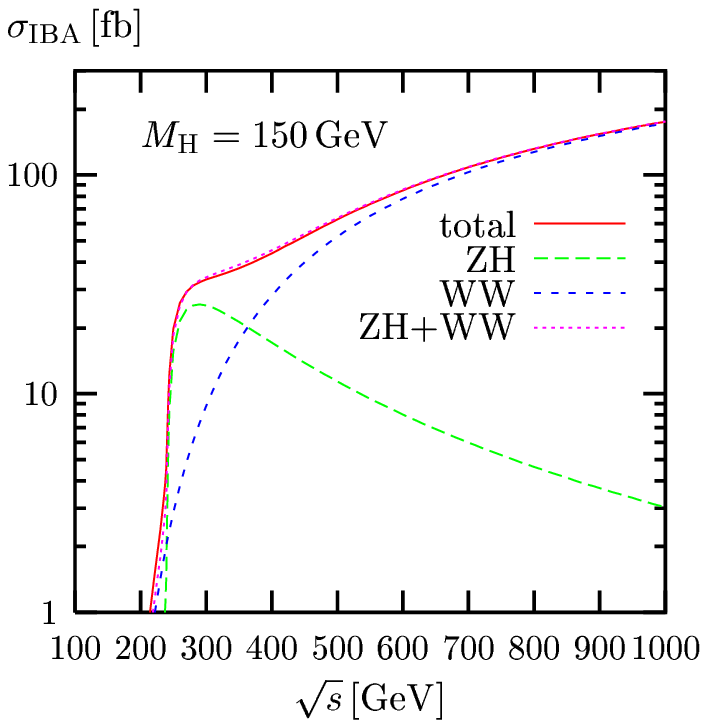}
\includegraphics[bb=40 415 250 625,width=.5\textwidth]{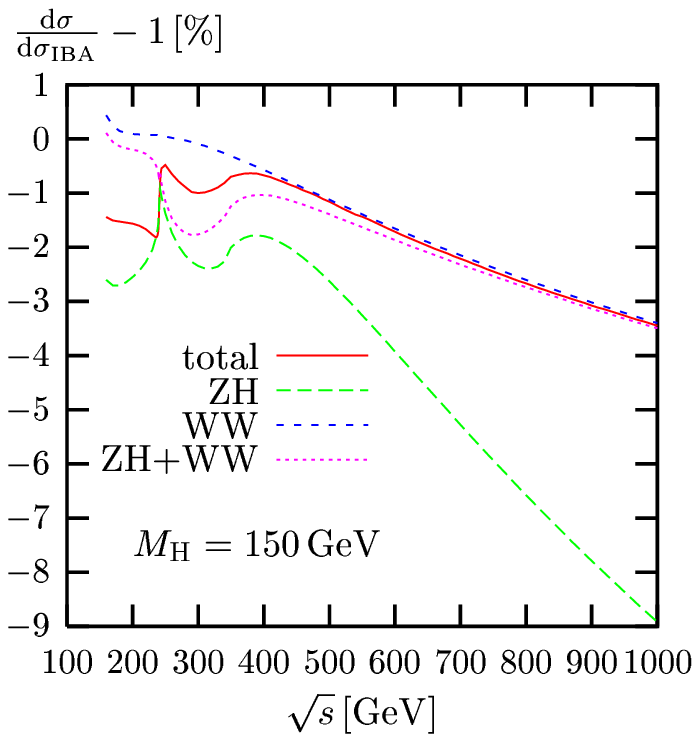}
}
\caption{Improved Born approximation (IBA) for the total cross sections 
and radiative corrections relative to the IBA for $\MH=150\GeV$}
\label{fig:totcs_iba}
\end{figure}

Figure~\ref{fig:totcs_iba} shows the total cross sections for the ZH
and WW channels, as well as their incoherent (``ZH+WW'') and coherent
(``total'') sums, in improved Born approximation (IBA) together with
the radiative corrections relative to the IBA as functions of the
centre-of-mass (CM) energy $\sqrt{s}$ for the fixed Higgs-boson mass
$\MH=150\GeV$.
Analogous results have been given in \citere{Denner:2003yg} for the
pure lowest-order cross sections $\sigma_{\mathrm{tree}}$ and the
corresponding relative corrections (see Figs.~1 and 2 there).  While
the absolute cross sections $\sigma_{\mathrm{IBA}}$ and
$\sigma_{\mathrm{tree}}$ look qualitatively similar, the relative
corrections normalized to the $\sigma_{\mathrm{IBA}}$ are
systematically smaller as compared to a normalization to the pure
lowest-order cross section. This is mainly due to the dominance of the
ISR corrections which are properly taken into account by the IBA.  In
the WW channel the IBA describes the corrected cross section within
1\% for CM energies up to $\sim 500\GeV$ and remains still good within
2--3\% up to $\sqrt{s}\lsim1\TeV$. At high CM energies the IBA misses
non-universal bosonic corrections, the size of which grows with energy
(compare Fig.~3 of \citere{Denner:2003yg}).  In the ZH channel the IBA
deviates from the corrected cross section by $\lsim3\%$ for
$\sqrt{s}\lsim500\GeV$, but even this reasonably good approximation
results from accidental compensations between fermionic and bosonic
non-ISR corrections, which are both of the order of 5--10\% but of
opposite sign (see Fig.~3 of \citere{Denner:2003yg}).  Above a CM
energy of $500\GeV$, where ZH production is suppressed, the IBA
becomes even worse, since the dominating Sudakov logarithms, such as
$\alpha\ln^2(s/\MW^2)$ are missing in the IBA.

\begin{figure}
\centerline{
\includegraphics[bb=40 415 250 625,width=.5\textwidth]{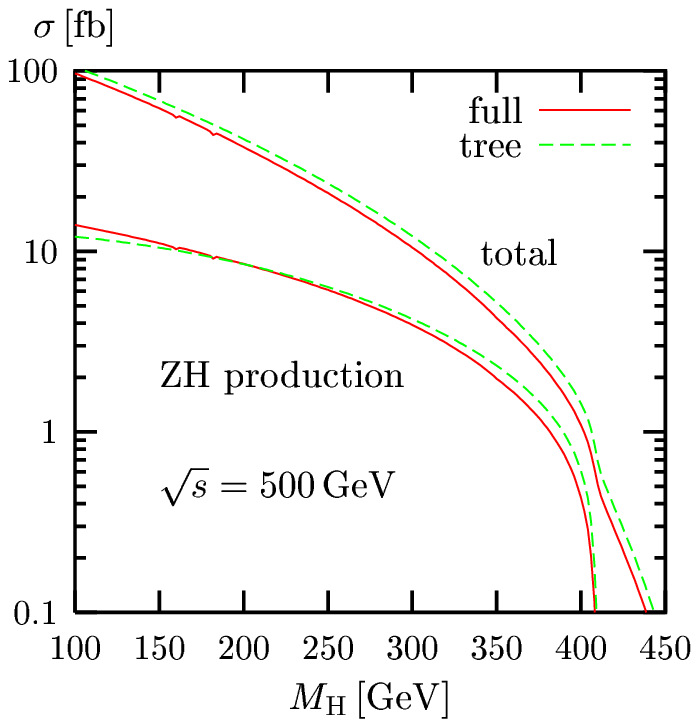} 
\includegraphics[bb=40 415 250 625,width=.5\textwidth]{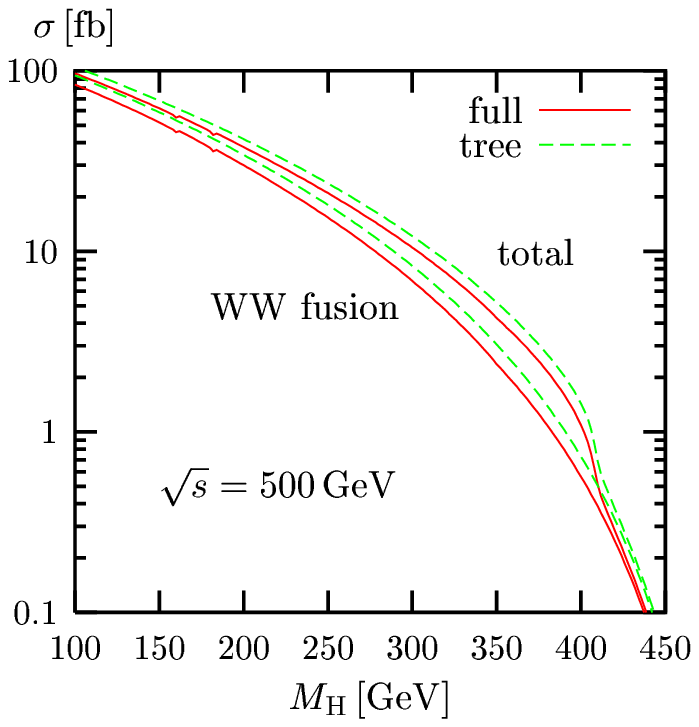} 
}
\vspace*{1em}
\centerline{
\includegraphics[bb=40 415 250 625,width=.5\textwidth]{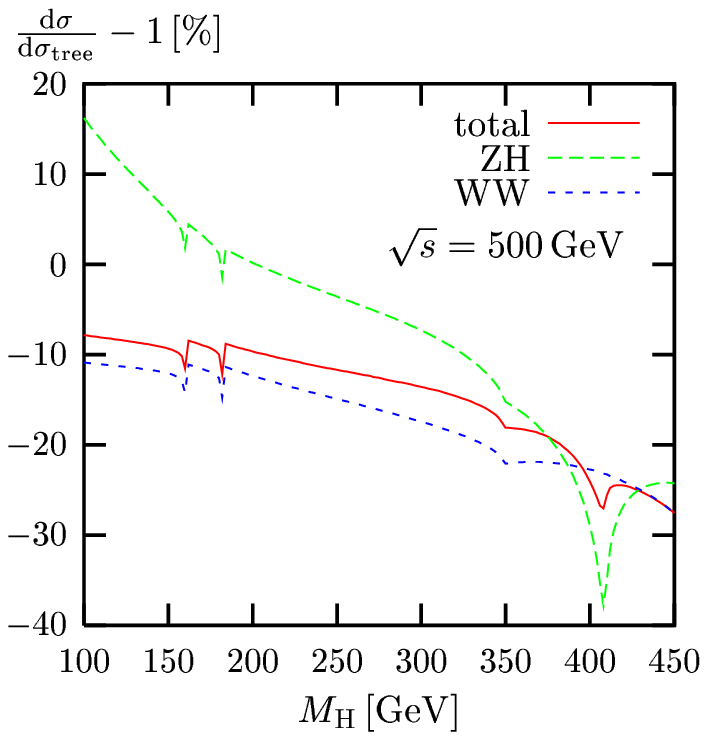}
\includegraphics[bb=40 415 250 625,width=.5\textwidth]{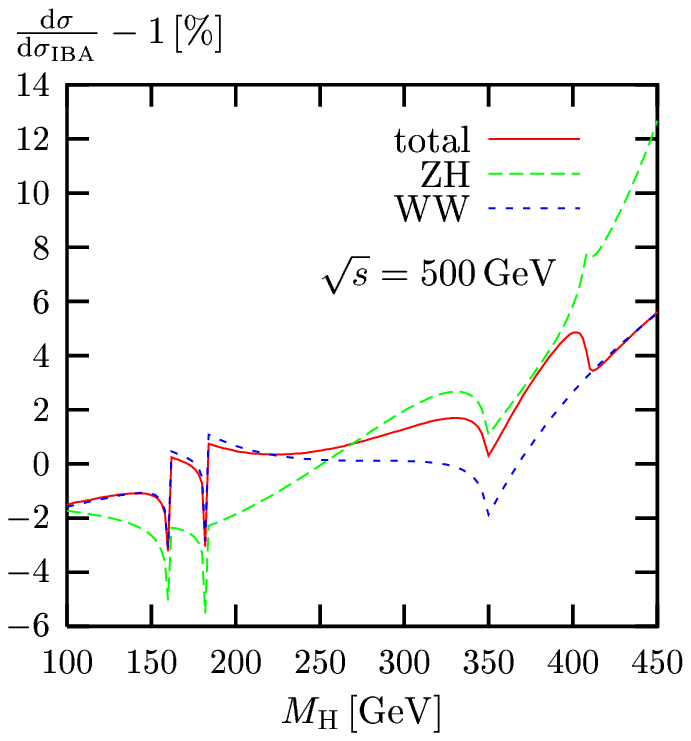}
}
\caption{Cross sections and relative corrections
as function of the Higgs-boson mass for $\sqrt{s}=500\GeV$}
\label{fig:totcs_mh}
\end{figure}

In \reffi{fig:totcs_mh} the cross sections and their corrections are
shown as functions of the Higgs-boson mass $\MH$ for the typical LC
energy of $\sqrt{s}=500\GeV$.
For this CM energy the contribution of WW fusion is much larger than
the one of ZH production for $\MH\lsim300\GeV$.  For
$\MH\gsim\sqrt{s}-\MZ\approx 409\GeV$ the ZH-production channel is
strongly suppressed since the Z boson cannot become resonant there.
The total cross section falls below $1\fb$ in this domain.  For
$\MH\lsim350\GeV$ the relative corrections to the total cross section
and to the WW-fusion channel are of the order $-10\%$ to $-20\%$ if
normalized to $\sigma_{\mathrm{tree}}$, but only at the level of 2\%
if normalized to $\sigma_{\mathrm{IBA}}$.  For the ZH-production
channel the reduction of the relative corrections with respect to
$\sigma_{\mathrm{IBA}}$ is similar.  The spikes at
$\MH=2\MW,2\MZ,2\Mt$ result from thresholds.  These singularities are
avoided if the finite widths of the unstable particles are taken into
account appropriately (see, for instance, \citere{Bhattacharya:1991gr}).

\subsection{Results on distributions}

\begin{figure}
\centerline{
\includegraphics[bb=40 415 250 625,width=.5\textwidth]{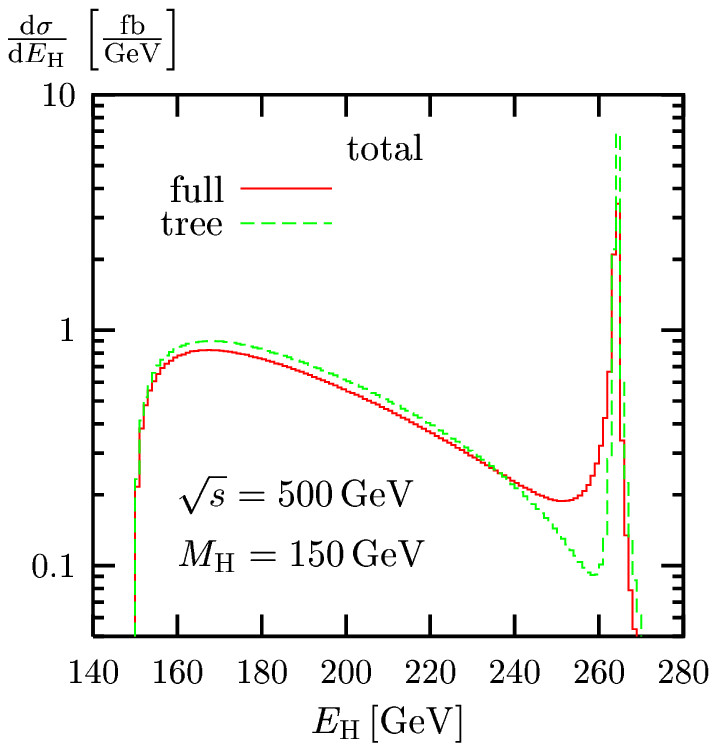}
\includegraphics[bb=40 415 250 625,width=.5\textwidth]{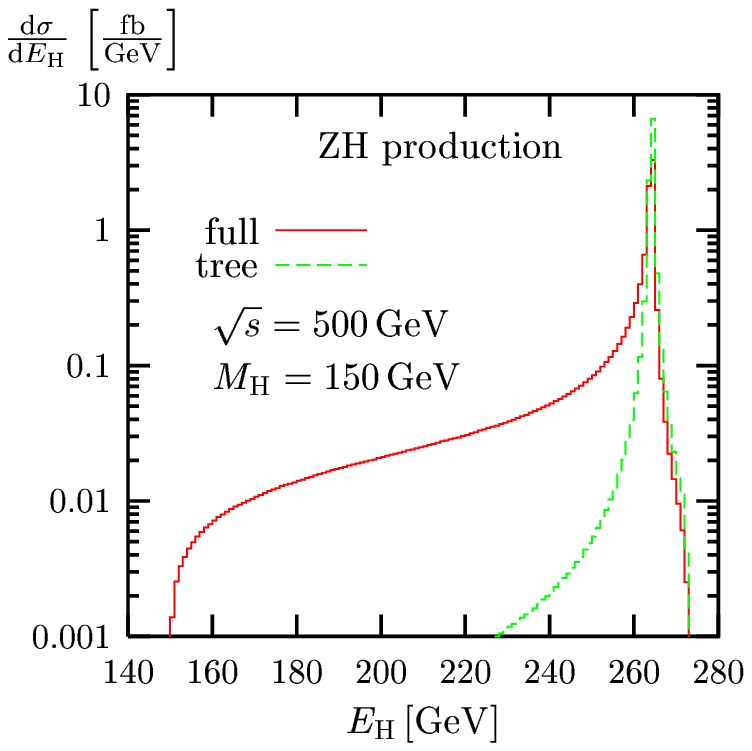}
}
{\unitlength 1cm
\begin{picture}(16,9)
\put(-1.6,-15.4){\includegraphics{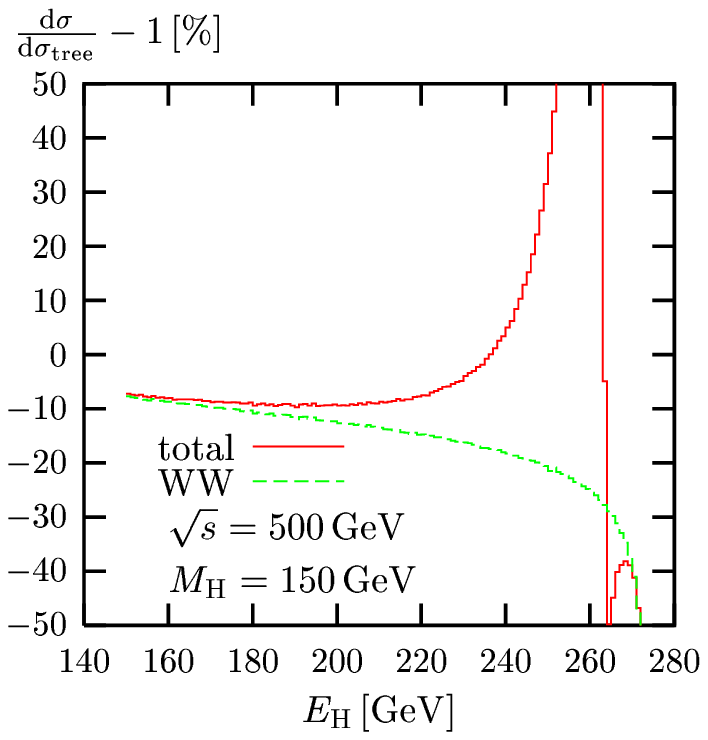}}
\put(0.7,-5.8){\includegraphics{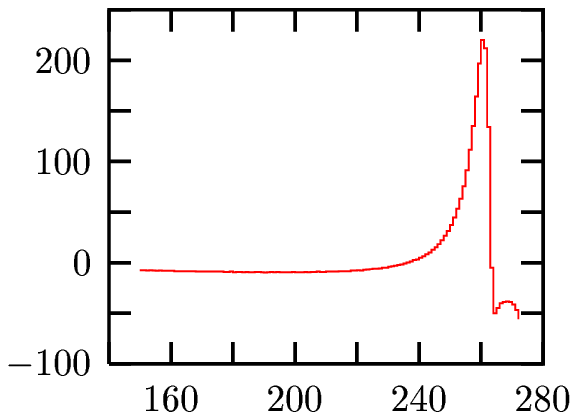}}
\put(6.5,-15.4){\includegraphics{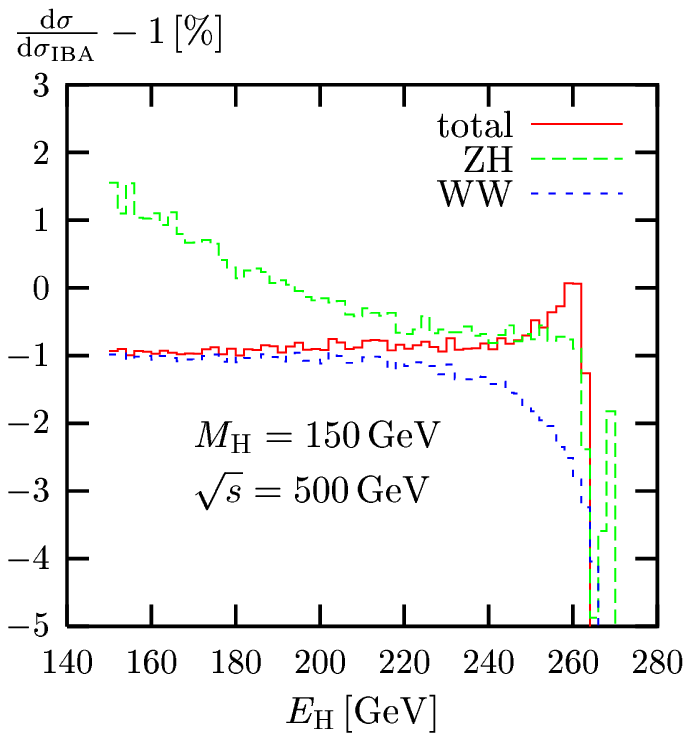}}
\end{picture}
}%
\caption{Distribution in the Higgs-boson energy $E_\PH$ and corresponding
relative radiative corrections for $\sqrt{s}=500\GeV$ and $\MH=150\GeV$}
\label{fig:EHdist}
\end{figure}

The distributions in the Higgs-boson energy $E_\PH$ 
(defined in the CM frame)
are depicted in
\reffi{fig:EHdist} for $\sqrt{s}=500\GeV$ and $\MH=150\GeV$.
The broad continuous distribution for $E_\PH\lsim250\GeV$ is almost
entirely due to WW fusion, while the resonance peak at
$E_\PH\sim265\GeV$ corresponds to the Higgs-boson energy in the
$2\to2$ kinematics of the ZH-production process with an on-shell
Z~boson. This explains the behaviour of the radiative corrections,
which are dominated by ISR.  Since the WW-fusion cross section rises
with energy, the continuous part receives negative corrections. For
the WW-fusion channel these are particularly large for large $E_\PH$
near the phase space boundary. On the other hand, the Z~peak is
reduced by ISR and produces a radiative tail for $E_\PH$ values below
the peak, since ISR effectively reduces the scattering energy of the
subsequent ZH-production process.  This radiative tail leads to
corrections of up to 220\%, as can be seen in the inset of the lower
left plot in \reffi{fig:EHdist}.  The relative corrections are reduced
drastically if normalized to the IBA, which takes care of the large
ISR effects.  For $E_\PH\lsim250\GeV$ the corrections to the total
improved-Born cross section vary only weakly with $E_\PH$.

\begin{figure}
\centerline{
\includegraphics[bb=40 415 250 625,width=.5\textwidth]{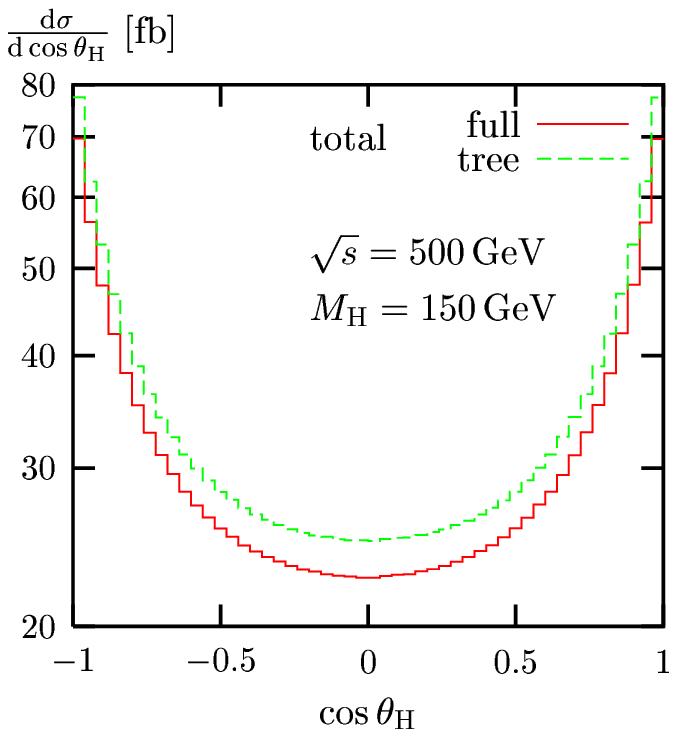}
\includegraphics[bb=40 415 250 625,width=.5\textwidth]{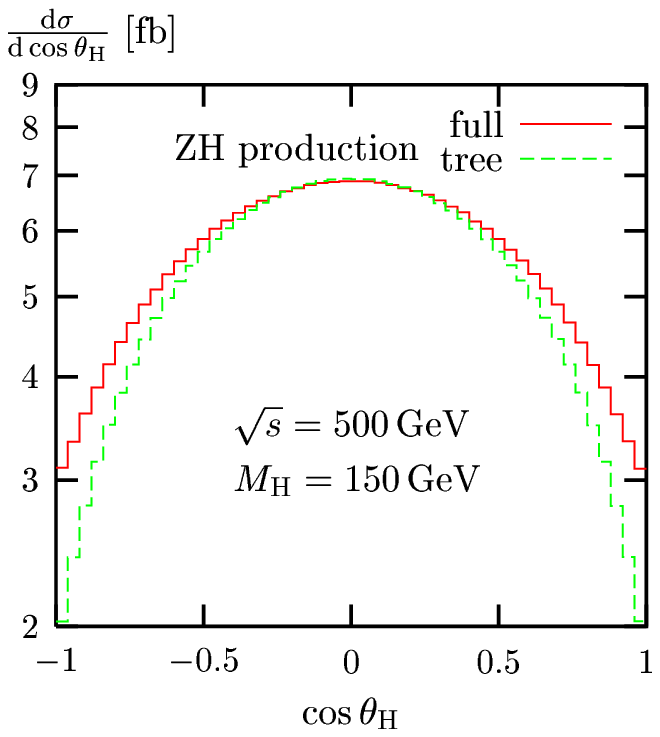}
}
\vspace*{1em}
\centerline{
\includegraphics[bb=40 415 250 625,width=.5\textwidth]{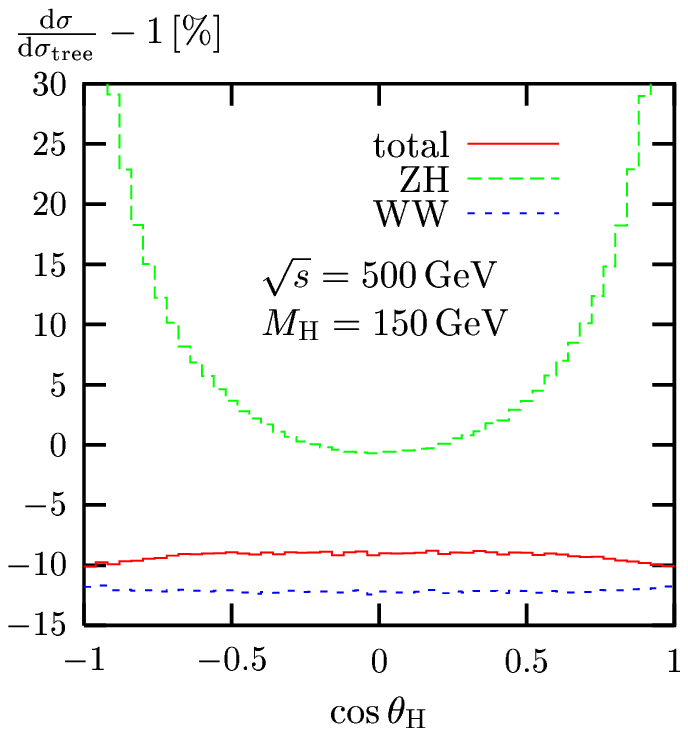}
\includegraphics[bb=40 415 250 625,width=.5\textwidth]{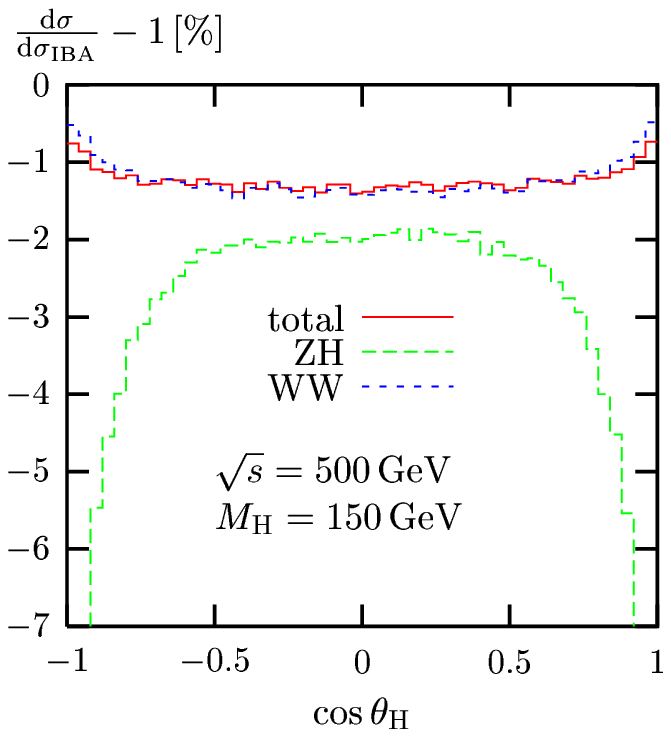}
}
\caption{Distribution in the cosine of the Higgs-boson production angle
$\theta_\PH$ and corresponding
relative radiative corrections for $\sqrt{s}=500\GeV$ and $\MH=150\GeV$}
\label{fig:cosHdist}
\end{figure}

Figure~\ref{fig:cosHdist} illustrates the distribution in the cosine of 
the Higgs-boson production angle $\theta_\PH$ 
(defined in the CM frame)
for $\sqrt{s}=500\GeV$ and 
$\MH=150\GeV$.
The peak behaviour in the very forward and backward directions is due
to the dominant WW contribution, while ZH production follows a shape
roughly proportional to $1-\cos^2\theta_\PH$.  The relative
corrections to the WW contribution depend only weakly on
$\cos\theta_\PH$ and reflect the same reduction after normalization to
the IBA as the integrated cross sections.  The relative corrections to
the ZH contribution become large in the forward and backward
directions, where the corresponding lowest-order cross section is
small.

Finally, we show the spectra of the photon energy $E_\gamma$ and the
photon polar angle $\cos\theta_\gamma$ (both defined in the CM frame)
of the radiative process $\eennh+\gamma$ for $\sqrt{s}=500\GeV$ and
$\MH=150\GeV$.
\begin{figure}
\centerline{
\includegraphics[bb=40 415 250 625,width=.5\textwidth]{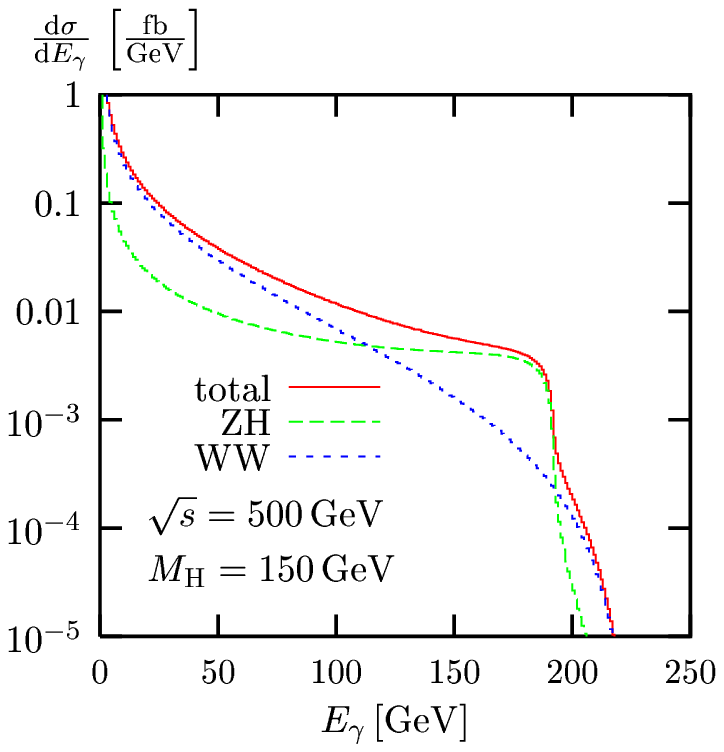}
\includegraphics[bb=40 415 250 625,width=.5\textwidth]{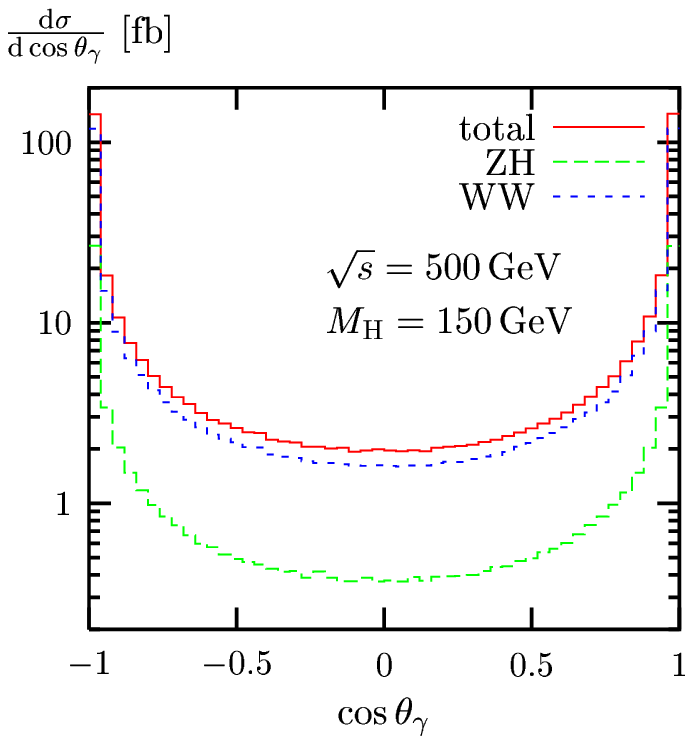}
}
\caption{Distribution in the photon energy $E_\gamma$ 
  and in the photon polar angle $\cos\theta_\gamma$ in the radiative
  process $\eennh+\gamma$ for $\sqrt{s}=500\GeV$ and $\MH=150\GeV$}
\label{fig:Egdist}
\end{figure}
In order to 
make the photon visible, we impose angular and energy cuts of
\beq
\theta(\gamma,\mathrm{beam}) > 1^\circ \qquad E_\gamma>0.1\GeV.
\eeq
For small photon energies the infrared pole, which is the same for the
WW and ZH channels, dominates the spectrum.  Thus, WW fusion dominates
for small $E_\gamma$, since the lowest-order cross section is much
larger for WW fusion than for ZH production.  For larger $E_\gamma$,
small-angle photon emission is dominant as some kind of remnant of the
collinear pole for forward and backward emission.  This means that
larger $E_\gamma$ reflect the lowest-order cross sections at smaller
energies.  Since the ZH-production cross section rises with decreasing
CM energies for $\sqrt{s}\lsim500\GeV$, but the WW-fusion cross
section falls off steeply, ZH production takes over the leading role
for $E_\gamma\gsim110\GeV$.

\subsection{Comparison to related work}

Adapting the input parameters and the parametrization of the
lowest-order matrix element to those used by Belanger et al.\ 
\cite{Belanger:2002me}, we reproduced the numbers for the total cross
section given in Table~2 of the first paper of
\citere{Belanger:2002me}.  Note that we switch off the ISR beyond
$\Oa$ in this comparison.  In \refta{ta:Boudjema} we list for each
Higgs-boson mass and the corresponding calculated
W-boson mass the results of \citere{Belanger:2002me}%
\footnote{According to F.~Boudjema, the numbers for the lowest-order
  cross section in Table~2 of \citere{Belanger:2002me} have
  integration errors of the order of $0.2\%$. Table \ref{ta:Boudjema}
  contains updated numbers obtained with increased statistics.}
together with our results. The numbers in parenthesis indicate the
errors in the last digits.
\begin{table}
$$\begin{array}{c@{\qquad}c@{\qquad}l@{\qquad}l@{\qquad}l@{\ \ }l}
\hline
\MH~[\mathrm{GeV}] & \MW~[\mathrm{GeV}] & \;\si_{\tree}~[\mathrm{fb}] &
\ \si~[\mathrm{fb}]~ & \si/\si_{\tree}-1~[\%] \\
\hline
150 & 80.3767     & 61.074(7)  & 60.99(7)  & -0.2     
& \mathrm{\citere{Belanger:2002me}}\\
    &             & 61.076(5)  & 60.80(2)  & -0.44(3) 
& \mathrm{this~work} \\
\hline
200 & 80.3571     & 37.294(4)  & 37.16(4)  & -0.4 
& \mathrm{\citere{Belanger:2002me}} \\
    &             & 37.293(3)  & 37.09(2)  & -0.56(4) 
& \mathrm{this~work} \\
\hline
250 & 80.3411     & 21.135(2)  & 20.63(2)  & -2.5 
& \mathrm{\citere{Belanger:2002me}} \\
    &             & 21.134(1)  & 20.60(1)  & -2.53(3) 
& \mathrm{this~work} \\
\hline
300 & 80.3275     & 10.758(1)  & 10.30(1)  & -4.2 
& \mathrm{\citere{Belanger:2002me}} \\
    &             & 10.7552(7) & 10.282(4) & -4.40(3) 
& \mathrm{this~work} \\
\hline
350 & 80.3158     & 4.6079(5)  & 4.184(4)  & -9.1 
& \mathrm{\citere{Belanger:2002me}} \\
    &             & 4.6077(2)  & 4.181(1)  & -9.27(3) 
& \mathrm{this~work} \\
\hline
\end{array}$$
\caption{Total cross section in lowest order and including the full
  $\Oa$ corrections and the relative corrections for
  $\sqrt{s}=500\GeV$ and various Higgs masses for the input parameter
  scheme of \citere{Belanger:2002me}${}^{\thefootnote}$}
\label{ta:Boudjema}
\end{table}
The agreement is good; the relative differences are with one exception
below $10^{-4}$ for the total lowest-order cross section and below
0.3\% for the corrected cross section.
The corrections relative to the lowest-order cross section agree
within 0.2\%.  This is of the order of the statistical error of
\citere{Belanger:2002me}, which is about 0.1\%.  Note that Belanger et
al.\ use $\alpha(0)$ to parametrize the lowest-order cross section.
As a consequence their relative corrections are shifted by $3\Delta
r\approx +9\%$ compared to those in the $\GF$-scheme.

We have also reproduced the $\cos\theta_\PH$ and $E_\PH$ distributions
in Figures 1 and 2 of the first paper of \citere{Belanger:2002me}.  We
found agreement within the accuracy of these figures.

When considering only fermion-loop corrections, we find agreement with
the calculations of \citeres{Eberl:2002xd,Hahn:2002gm}, once the
appropriate renormalization and input-parameter schemes are adopted.
For more details on this comparison we refer to
\citere{Denner:2003yg}.

\section{Summary}
\label{se:sum}

We have presented a calculation of the complete electroweak ${\cal
  O}(\alpha)$ radiative corrections to the single Higgs-boson
production process $\Pep\Pem\to \nu\bar\nu\PH$ in the electroweak
Standard Model.  For $\Pep\Pem\to\nu_\Pe\bar\nu_\Pe\PH$, where
$\PZ\PH$ production and W-boson fusion contribute, both production
channels are added coherently. Two methods for the treatment of the
finite Z-boson width have been introduced. We have taken special care
to treat the contributions from 5-point tensor integrals in a
numerically stable way. The complete single-hard-photon matrix
elements have been taken into account, where soft and collinear
singularities are treated both in the subtraction and the phase-space
slicing methods.  Higher-order ISR corrections have been included in
the structure-function approach.  The phase-space integration is
performed with Monte-Carlo techniques.

We find that the electroweak corrections are of the order of $-10\%$
and are dominated by ISR corrections if the lowest-order matrix
element is parametrized with the Fermi constant $\GF$.  The non-ISR
corrections are at the level of a few per cent in the $\GF$-scheme,
but are of the order of 10\% in other schemes.  At high energies,
where the WW-fusion channel dominates, the electroweak corrections
depend only weakly on the energy and the production angle of the
Higgs-boson.

Although the sum of ISR and non-ISR is accidentally small in the
$\alpha(0)$-scheme, the $\GF$-scheme is nevertheless preferable for
the following reasons. In contrast to the $\alpha(0)$ scheme, it does
not suffer from uncertainties arising from the hadronic vacuum
polarization at low energies. Since the cancellation between ISR and
non-ISR corrections in the $\alpha(0)$ scheme at one loop is
accidental, it cannot be expected that it still holds in higher
orders. One the other hand, the $\GF$-scheme resums the leading
universal corrections associated with the running of the
electromagnetic coupling and the universal corrections proportional to
the square of the top mass.

We have shown that the corrections to the WW-fusion channel can be
described by a simple improved Born approximation within an accuracy
of typically 1\% (3\%) for CM energies below $500\GeV$ ($1\TeV$).  For
the ZH-production channel the improved Born approximation, which is
simply based on ISR, approximates the corrected cross section within
3\% up to CM energies of about $500\GeV$, but becomes worse at higher
energies. In summary, the approximation can be used to include
radiative corrections at the qualitative level, but a precision
analysis will require the inclusion of the full ${\cal O}(\alpha)$
correction.

\appendix
\section*{Appendix}
\renewcommand{\theequation}{A.\arabic{equation}}

\section*{Standard matrix elements}
\label{app:sme}

The four-dimensionality of space--time implies that the SME
$\Msme^{\ZH,\si}_{i}$ and $\Msme^{\WW}_{i}$ introduced in
\refse{se:convs} are not all independent; there are linear relations
among them.

A simple way to derive the relations with real coefficients is
provided by the following trick. In four dimensions the metric tensor
can be decomposed in terms of four independent orthonormal
four-vectors $n_l$,
\beq
g^{\al\be} = n_0^\al n_0^\be - \sum_{l=1}^3 n_l^\al n_l^\be,
\label{eq:gdecomp2}
\eeq
where $n_k\cdot n_l = g_{kl}$.
Two convenient choices ($j=1,2$) for the vectors $n_l$ are given by
\beqar
n_0^\al &=& \frac{1}{\sqrt{s}}(p_1+p_2)^\al, \qquad
n_1^\al = \frac{1}{\sqrt{s}}(p_1-p_2)^\al, 
\nn\\
n_2^\al &=& \sqrt{\frac{s}{t_{1j}t_{2j}}}\left( k_j^\al
        +\frac{t_{2j}}{s}p_1^\al + \frac{t_{1j}}{s}p_2^\al \right), \quad
\nn\\
n_3^\al &=& \epsilon^{\al\be\ga\de} n_{0,\be}n_{1,\ga}n_{2,\de}=
-\frac{2}{\sqrt{st_{1j}t_{2j}}} \epsilon^{\al\be\ga\de}
        p_{1,\be}p_{2,\ga}k_{j,\de},
\eeqar
with $\epsilon^{\al\be\ga\de}$ ($\epsilon^{0123}=+1$) denoting the
totally antisymmetric tensor. Inserting this decomposition for both
$j=1,2$ in all contractions between different Dirac chains according
to $\Ga^{\Pe\Pe,\si}_{\al}\Ga^{\nu\nu,\al} =
\Ga^{\Pe\Pe,\si}_{\al}g^{\al\be}\Ga^{\nu\nu}_{\be}$, etc., and
subsequently using the Dirac equation and the Chisholm identity
\beq
 \ri\epsilon^{\al\be\ga\de}\ga_\de\ga_5=
\ga^\al \ga^\be \ga^\ga - g^{\al\be}\ga^\ga + g^{\al\ga}\ga^\be
- g^{\be\ga}\ga^\al,
\label{eq:chisholm}
\eeq
reduces all ZH SME $\Msme^{\ZH,\si}_{i}$ to four and all WW SME
$\Msme^{\WW}_{i}$ to two SME.

Further relations with complex coefficients result from the direct
application of the Chisholm identity \refeq{eq:chisholm} to structures
like $\Ga^{\nu\nu}_{k_1 p_1 k_2}$ and subsequently using the
decomposition
\beq
g^{\rho\si} = \sum_{i,j=1}^4 \left(Z^{-1}\right)_{ij} 2p_i^\rho p_j^\si,
\qquad Z_{ij} = 2p_i p_j,
\qquad p_3=k_1, \quad p_4=k_2,
\eeq
to separate all contractions between $\eps$ tensors and Dirac chains
via $\Ga^{\nu\nu}_{\de} \eps^{\al\be\ga\de} = \Ga^{\nu\nu}_{\rho}
g^{\rho\si} \eps^{\al\be\ga}_{\phantom{\al\be\ga}\si}$.  For
$\Ga^{\nu\nu}_{k_1 p_1 k_2}$, in particular, this leads to
\beq
\Ga^{\nu\nu}_{\de} \eps^{k_1 p_1 k_2\de} 
= \Ga^{\nu\nu}_{\rho} g^{\rho\si} \eps^{k_1 p_1 k_2}_{\phantom{k_1 p_1 k_2}\si}
= -2X \left[ \Ga^{\nu\nu}_{p_1} \left(Z^{-1}\right)_{12} 
+ \Ga^{\nu\nu}_{p_2} \left(Z^{-1}\right)_{22} \right]
\eeq
with
\beq
X = \eps^{p_1 p_2 k_1 k_2}
= \eps^{\mu\nu\rho\si} p_{1,\mu} p_{2,\nu} k_{1,\rho} k_{2,\si}.
\eeq
In the inverse matrix $\left(Z^{-1}\right)$, the determinant $\det(Z)$
occurs, which can be identified with $\det(Z)=-16X^2$.

Altogether, the linear relations reduce the set of ZH SME to two and
the set of WW SME to one SME.  Explicitly, the relations read
\beq
\Msme^{\ZH,\pm}_i = r_i^{\ZH,\pm} \Msme^{\ZH,+}_1,\qquad
\Msme^{\WW}_i =   r_i^{\WW}\Msme^{\WW}_1,
\qquad i=2,\ldots13,
\eeq
with
\beq
\begin{array}[b]{lllllllll}
r^{\ZH,+}_2 &=& s ,&
r^{\ZH,-}_2 &=& 0,&
r^{\WW}_2 &=& - t_{11} ,\\
r^{\ZH,+}_3 &=& 0,&
r^{\ZH,-}_3 &=& s_{12} ,&
r^{\WW}_3 &=& 0,\\
r^{\ZH,+}_4 &=& 0,&
r^{\ZH,-}_4 &=& \frac{1}{2} C^* , &
r^{\WW}_4 &=& \frac{1}{2} C ,\\
r^{\ZH,+}_5 &=& \frac{1}{4 t_{22}} A ,\qquad& 
r^{\ZH,-}_5 &=& -\frac{t_{11}}{2} ,&
r^{\WW}_5 &=& \frac{s}{2} ,\\
r^{\ZH,+}_6 &=& - \frac{t_{21}}{2} ,&
r^{\ZH,-}_6 &=& \frac{1}{4 t_{12}} A^*  ,\qquad&
r^{\WW}_6 &=& -\frac{1}{4t_{12}} B^* ,\\
r^{\ZH,+}_7 &=& - \frac{t_{12}}{2} , &
r^{\ZH,-}_7 &=& \frac{1}{4 t_{21}} A  ,&
r^{\WW}_7 &=& -\frac{1}{4 t_{21}} B  ,\\
r^{\ZH,+}_8 &=& \frac{1}{4 t_{11}} A^* , &
r^{\ZH,-}_8 &=& -\frac{t_{22}}{2} , &
r^{\WW}_8 &=&  \frac{s_{12}}{2} ,\\
r^{\ZH,+}_9 &=& \frac{1}{t_{22}}  A , &
r^{\ZH,-}_9 &=& -4 t_{11} ,&
r^{\WW}_9 &=& 4 s ,\\
r^{\ZH,+}_{10} &=& -2 t_{21} ,&
r^{\ZH,-}_{10} &=& \frac{2}{t_{12}} A^* , &
r^{\WW}_{10} &=& -\frac{2}{t_{12}} B^*  ,\\
r^{\ZH,+}_{11} &=& -2 t_{12} ,&
r^{\ZH,-}_{11} &=& -4 t_{12} ,&
r^{\WW}_{11} &=& -4 t_{12} ,\\
r^{\ZH,+}_{12} &=& \frac{1}{t_{11}} A^* &
r^{\ZH,-}_{12} &=& -4 t_{22} ,&
r^{\WW}_{12} &=& 4 s_{12} ,\\
r^{\ZH,+}_{13} &=& 4  ,& 
r^{\ZH,-}_{13} &=& 16 ,&
r^{\WW}_{13} &=& 16
\end{array}
\eeq
and
\beqar
A &=& s s_{12} - t_{12} t_{21} - t_{11} t_{22} + 4 \ri X,\nl
B &=& s s_{12} + t_{12} t_{21} - t_{11} t_{22} + 4 \ri X,\nl  
C &=& s s_{12} - t_{12} t_{21} + t_{11} t_{22} + 4 \ri X.
\eeqar                                  

Finally, there are Fierz identities relating the ZH and WW spinor
chains, supplementing the above list of relations by
\beq
\Msme^{\WW}_1 = -\Msme^{\ZH,-}_1,
\eeq
so that all SME can be expressed in terms of $\Msme^{\ZH,\pm}_1$.
In terms of Weyl--van der Waerden spinor products, which have been 
defined in \refse{se:bremsme} (see also \citere{Dittmaier:1999nn}),
these SME read
\beq
\Msme^{\ZH,+}_1 = 2\langle p_2 k_2 \rangle^* \langle p_1 k_1\rangle, 
\qquad
\Msme^{\ZH,-}_1 = 2\langle p_1 k_2 \rangle^* \langle p_2 k_1\rangle.
\eeq

Since the lowest-order matrix elements for right- and left-handed
electrons are proportional to $\Msme^{\ZH,+}_1$ and $\Msme^{\ZH,-}_1$,
respectively, the whole virtual one-loop contributions to the squared
matrix element are of the form
\beq
\Re\left\{\M_1^\si\left(\M_0^\si\right)^*\right\} = 
\left[ f_1^\si(s,t_{ij},s_{ij}) + f_2^\si(s,t_{ij},s_{ij})X\right] 
|\Msme_1^{\ZH,\si}|^2,
\eeq
where $f^\si_i$ are functions of scalar products of the external
momenta.  Note that the $f^\si_i$ and $|\Msme_1^{\ZH,\si}|^2$ are
invariant under the reflection of all outgoing momenta ${\bf k}_j$ on
the plane spanned by the beam axis and the Higgs-boson momentum ${\bf
  k}_3$, while $X$ changes its sign under this reflection.
Consequently, the contribution proportional to $X$ drops out after
integrating over the momenta of the final-state neutrinos, which are
not observable.

\section*{Acknowledgement}

We thank the authors of
\citeres{Belanger:2002me,Eberl:2002xd,Hahn:2002gm} for further
information about their results and, in particular, F.~Boudjema for
sending us more precise numbers.  This work was supported in part by
the Swiss Bundesamt f\"ur Bildung und Wissenschaft and by the European
Union under contract HPRN-CT-2000-00149.

\end{document}